\documentclass[journal]{IEEEtran}
\usepackage{amsmath}
\usepackage{amssymb}
\usepackage{amsthm,amsmath,amssymb}
\usepackage{mathrsfs}
\usepackage{algorithm}  
\usepackage{algpseudocode}   

\usepackage{graphicx}
\graphicspath{{figures/}}

\usepackage{color}
\usepackage{graphicx}
\usepackage{float}
\usepackage{subfig}
\usepackage{algorithmicx}
\usepackage{algorithm}
\usepackage{algpseudocode}

\usepackage{bm}

\usepackage{cite}

\usepackage[colorlinks=true,
linkcolor=blue,   
citecolor=blue,   
urlcolor=blue,    
anchorcolor=blue]{hyperref}

\ifCLASSINFOpdf
\else
\fi
\begin{document}
	
	\title{Spatial Crowdsourcing-based Task Allocation for UAV-assisted Maritime Data Collection}
	
	
	\author{Xiaoling Han,
		Bin Lin*,~\IEEEmembership{Senior Member,~IEEE,}
		Zhenyu Na,~\IEEEmembership{Member,~IEEE,}
		Bowen Li,
		Chaoyue Zhang,~\IEEEmembership{Student Member,~IEEE,}
		and~Ran Zhang,~\IEEEmembership{Senior Member,~IEEE}
		
	\thanks{
			Copyright (c) 20xx IEEE. Personal use of this material is permitted. However, permission to use this material for any other purposes must be obtained from the IEEE by sending a request to pubs-permissions@ieee.org.
	}
	
	\thanks{
	    	Manuscript received XXX; revised XXX; accepted XXX. Date of publication XXX; date of current version XXX. 
	}
		
	\thanks{
			The work was supported by the National Natural Science Foundation of China (No. 62371085, No. 51939001). (Corresponding author: Bin Lin.)
	}
		
	\thanks{
			Xiaoling Han, Bin Lin, Zhenyu Na, and Chaoyue Zhang are with the Department of Information Science and Technology of Dalian Maritime University, Dalian, China. E-mail: xiaolinghan@dlmu.edu.cn, binlin@dlmu.edu.cn, nazhenyu@dlmu.edu.cn, and zcy\_11335577@163.com.
			
			Bowen Li is with theDepartment of Artificial Intelligence and Software, Liaoning Petrochemical University, Fushun, China. E-mail: libowen@dlmu.edu.cn.
   
			Ran Zhang is with the Department of Electrical and Computer Engineering, University of North Carolina at Charlotte, Charlotte, NC 28223-0001 USA. E-mail: rzhang8@charlotte.edu.
			
			This manuscript has been accepted by IEEE Transactions on Vehicular Technology, DOI: 10.1109/TVT.2024.3483890.
	}
	}

	\markboth{Journal of \LaTeX\ Class Files,~Vol.~14, No.~8, August~2015}%
	{Shell \MakeLowercase{\textit{et al.}}: Bare Demo of IEEEtran.cls for IEEE Journals}
	%

	\maketitle
	
	\begin{abstract}
		Driven by the unceasing development of maritime services, tasks of unmanned aerial vehicle (UAV)-assisted maritime data collection (MDC) are becoming increasingly diverse, complex and personalized. As a result, effective task allocation for MDC is becoming increasingly critical. In this work, integrating the concept of spatial crowdsourcing (SC), we develop an SC-based MDC network model and investigate the task allocation problem for UAV-assisted MDC. In variable maritime service scenarios, tasks are allocated to UAVs based on the spatial and temporal requirements of the tasks, as well as the mobility of the UAVs. To address this problem, we design an SC-based task allocation algorithm for the MDC (SC-MDC-TA). The quality estimation is utilized to assess and regulate task execution quality by evaluating signal to interference plus noise ratio and the UAV energy consumption. The reverse auction is employed to potentially reduce the task waiting time as much as possible while ensuring timely completion. Additionally, we establish typical task allocation scenarios based on maritime service requirements indicated by electronic navigational charts. Simulation results demonstrate that the proposed SC-MDC-TA algorithm effectively allocates tasks for various MDC scenarios. Furthermore, compared to the benchmark, the SC-MDC-TA algorithm can also reduce the task completion time and lower the UAV energy consumption.
	\end{abstract}
	

	%
	\IEEEpeerreviewmaketitle
	
    \section{Introduction}

        \IEEEPARstart{D}{riven} by the continuous development of maritime services such as marine resources exploration, reconnaissance and surveillance, anti-submarine, marine tourism, marine transportation and emergency collection, tasks of unmanned aerial vehicle (UAV)-assisted maritime data collection (MDC) are becoming increasingly diverse, complex and personalized \cite{1,2,3}. Providing efficient and reliable data transmission, the MDC has gradually become one of the main subjects in maritime research field and promoted the application of the internet of things (IoT) in smart ocean \cite{4}. In maritime networks, low data transmission rates, limited coverage and lack of flexibility bring significant challenges to the MDC task execution \cite{5}. Therefore, it is indispensable to develop schemes for efficient MDC task allocation with considering different requirements in various maritime service scenarios.	
        
        There have been quite a few existing works for the MDC by analyzing the service requirements. In different maritime service scenarios, there are different requirements on the MDC in terms of mobility, bandwidth, delay and energy consumption \cite{6,7}. For instance, in the scenario of the positioning, navigation and emergency communication for maritime users, data transmission efficiency is prioritized \cite{8}, while in the scenario of the audio and video transmission for maritime users, large data transmission needs to be considered \cite{9}. Lyu \textit{et al.} \cite{10} proposed a fast UAV trajectory planning algorithm based on Fermat point theory in the maritime IoT system. By deploying UAVs, appropriate hovering points were selected to improve the channel conditions and increase the amount of data collected by UAVs. Further, Li \textit{et al.} \cite{11} investigated the joint UAV trajectory and transmit power optimization problem to enhance the coverage of a space-ground hybrid maritime communications network. In order to better observe the ocean independently, Leonard \textit{et al.} \cite{12} derived optimal paths for the mobile sensor network to achieve the best data collection. Shen \textit{et al.} \cite{41} studied data collection for massive machine-type communications networks enabled by UAV stations moving in the air. However, the combined consideration of the spatial and temporal requirements of tasks and the mobility of UAVs often becomes a key factor limiting performance in MDC task allocation.
        
        Different from the above task allocation methods, many works have proposed spatial crowdsourcing (SC) to allocate tasks \cite{29,30,31}. SC is a new paradigm of crowdsourcing platforms that can provide higher task completion rates, making it highly promising for practical applications. Workers are invited to move to designated target positions and perform tasks according to the server instructions, leading to higher task completion rates, and reduced task execution time and costs \cite{13}. Xiong \textit{et al.} \cite{14} defined a new spatial and temporal coverage metric, which jointly considers the proportion of subareas covered by sensors and the sensor number in each covered subarea. With the metric, they proposed a swarm sensing task allocation model to collect data and exploited large-scale real-world datasets for model evaluation. Wang \textit{et al.} \cite{20} divided the overly complex mobile crowdsourcing tasks into many simpler sub-tasks and optimized the scheduling considering the reliability of workers to jointly minimize the task completion time and the overall idle time.
        
        Employing SC to allocate tasks can comprehensively consider the task diversity and better optimize the task allocation scheme \cite{21,22}. Task requesters can set the task deadlines and rewards based on the task urgency and complexity. Considering the problems of quality estimation and monetary incentives, Yang \textit{et al.} proposed a quality-based truth estimation and surplus sharing method \cite{15}. Considering the strategic behaviors of mobile users, Zheng \textit{et al.} modeled the weighted coverage maximization under different coverage requirements in mobile crowdsensing as budget-limited reverse auctions \cite{16}. Wang \textit{et al.} designed a threshold-based online task allocation mechanism, which can handle the multiple task allocation to a worker \cite{18}. According to the position of workers and tasks, tasks were allocated in real time and the path planning was considered to make the task allocation more reasonable. Chen \textit{et al.} studied the problem of vehicle crowdsourcing and proposed a time-sensitive incentive mechanism to utilize the powerful onboard capability to perform various tasks in smart cities \cite{17}. Wu \textit{et al.} studied task allocation in edge computing environment and proposed an online incentive mechanism to maximize the utility of platform and crowd workers \cite{19}.
        
        Most existing works on SC have been conducted in terrestrial environments, with very few studies integrating SC into maritime environments. Unlike the SC applications in terrestrial environments (e.g., taxi hailing where there are a large number of vehicles and replenishing energy is convenient), the UAV number in marine environments is small and energy replenishing is inconvenient. As a result, task balance among all UAVs to improve their overall utilization subject to the limited battery capacities become new critical challenges. 
        
        This work focuses on the SC-based task allocation for the MDC, optimizing the task allocation considering two decisive factors: i) the spatial and temporal constraints of the MDC tasks, and ii) the mobility and energy limitations of UAVs. In addition, the complex marine environment has a significant impact on the reliability of UAV communication, such as communication jitter. Therefore, we choose large-size rotary-wing UAVs. They can provide powerful and stable communication capabilities, enabling them to withstand adverse environmental conditions. With these advantages, UAVs can rapidly collect and transmit data. Specifically, the main contributions of this work are summarized as follows.
	
	\begin{figure*}
		\centering
		\includegraphics[width=1\linewidth]{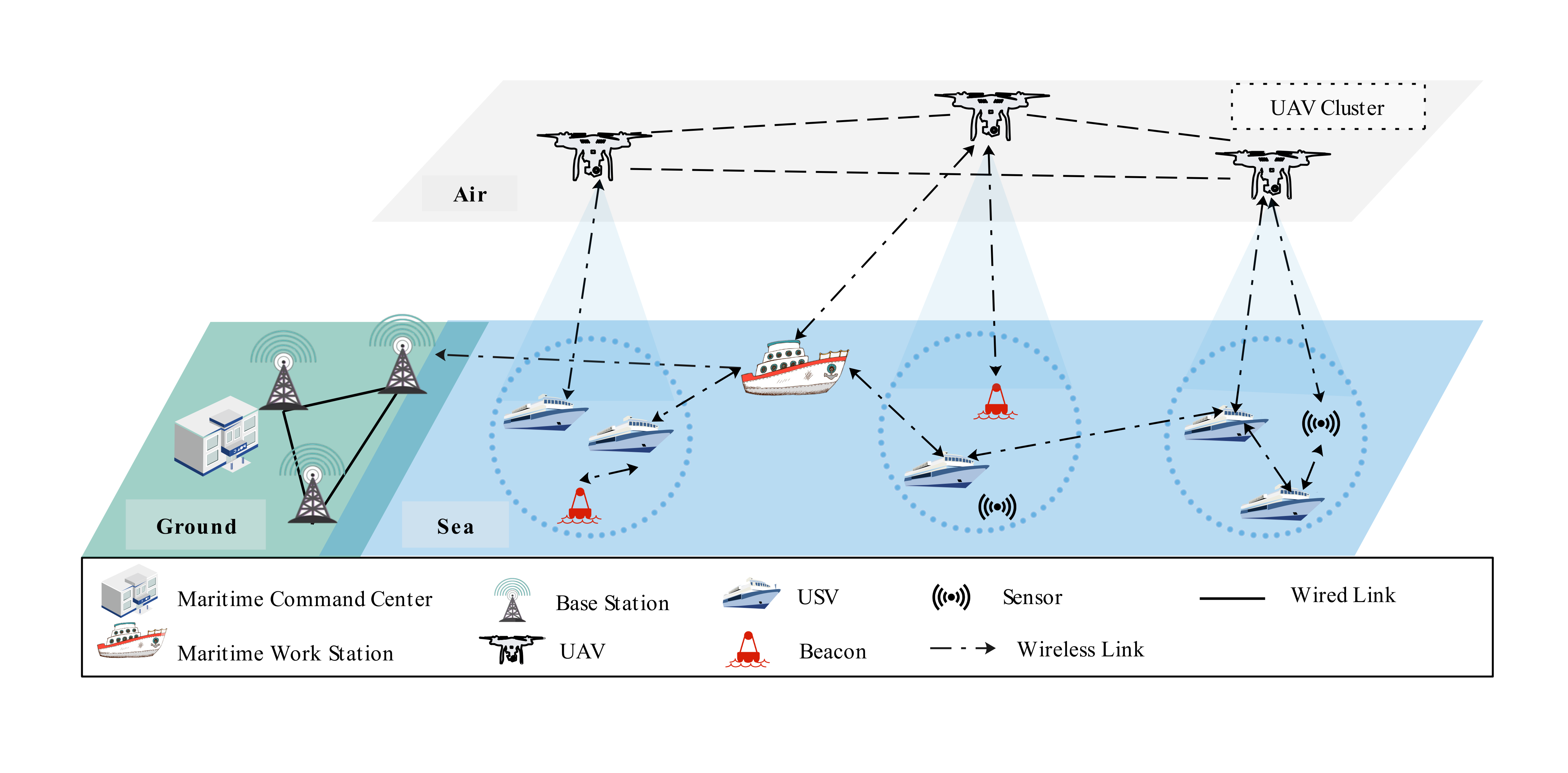}
		\caption{The proposed SC-based MDC network model.}\label{MDCNetwork}
	\end{figure*}

	\begin{enumerate}
		\item	An SC-based MDC network model is developed. The model integrates the maritime command center on the ground, UAVs in the air, the maritime work station and a set of agents at sea, including unmanned surface vehicles (USVs), beacons and sensors. Agents at sea publish tasks to the maritime work station requiring specific maritime services. The maritime work station allocates tasks to UAVs, and UAVs execute tasks accordingly.
		\item	An MDC task allocation problem is formulated. The problem aims to minimize the average task completion time by allocating tasks to UAVs. In the formulation, we divide the area covered by the UAV, assess the maritime channel conditions with signal to interference plus noise ratio (SINR), and calculate the UAV energy consumption and the task waiting time.
		\item	An SC-MDC-TA algorithm is proposed to solve the optimization problem. Specifically, the quality estimation ensures the task execution quality, while the reverse auction ensures the minimum task waiting time under the condition that tasks are executed within the validity period. 
		\item	Extensive simulations are conducted to verify the efficacy of the proposed algorithm. According to the electronic navigational charts, four scenarios are established. Compared with the benchmark, the proposed SC-MDC-TA algorithm can shorten the task completion time and reduce the UAV energy consumption.
	\end{enumerate}
	
        The remainder of this paper is organized as follows. Section II presents the SC-based MDC network model and then the corresponding optimization problem is formulated. Then, Section III introduces the details of the proposed algorithm. Section IV shows the simulations and performance evaluations. Finally, section V concludes this work.

   \section{System Model and Problem Formulation}

        In this section, the MDC is introduced. We describe the SC-based MDC network model and describe the MDC task allocation procedure. Then, we focus on the SC-based task allocation problem and formulate it.
	
   \subsection{Maritime Data Collection}
 
        Agents on the sea as the monitoring terminals will generate various maritime data. However, due to the complex maritime environment, it is difficult to lay out the communication infrastructure to transmit data directly as in terrestrial scenarios. At the same time, due to the limited energy of monitoring terminals and the high cost of satellite access, it is infeasible to upload large amounts of maritime data to satellites. Therefore, UAV-assisted MDC is a promising solutions which can effectively save communication cost and improve system efficiency \cite{24}.

        The MDC requires UAVs to perform tasks in designated area in a specific time. The MDC network's operation mechanism is influenced by a multitude of factors  \cite{26}. These include the limited resources and energy of agents, performance heterogeneity of physical devices, and the limited communication coverage. Generally speaking, the generated maritime data is temporal and spatial, and the UAV mobility can well cope with this situation well. The maritime work station allocates tasks to UAVs to facilitate efficient MDC.

   \subsection{Network Model}

        Integrating the concept of SC, we propose an SC-based MDC network model, as shown in Fig. \ref{MDCNetwork}. 

        The ground sub-network includes coastal base stations and the maritime command center. For the complex marine business, the maritime command center manages maritime networks and provides maritime services. Prior to MDC tasks, the maritime command center dispatches a number of UAVs, USVs and the maritime work station to conduct maritime operations and monitor the sea area. After MDC, the maritime command center analyzes the data and makes manual decisions.

        The air sub-network consists of multiple UAVs, which are responsible for executing tasks. Corresponding to the workers in the SC, UAVs go to designated positions to execute specific MDC tasks according to the received instructions. Because the UAV has a large capacity, data is not unloaded during the task execution. After all tasks are completed, the data is brought back to the origin of UAVs to unload. 

        In addition, the sea sub-network consists of two aspects. 

        One is the agents that publish tasks, including USVs, beacons and sensors. Corresponding to the task requesters in SC, during daily operation on the sea surface, they collect important data such as water temperature, weather and traffic flow, and then wait for the generated voice, image, and video data to be collected. Among them, the beacons and sensors are deployed in advance for the sea area. In addition to collecting maritime data, USVs can also act as a relay node. 

        The other is the maritime work station, corresponding to the SC server. It is critical for the maritime work station to carefully plan task allocation to maximize the utilization of UAV resources and improve the task execution efficiency \cite{25}. The maritime work station is not only portable for receiving data and issuing commands, but also equipped with robust computing capabilities. It analyzes the spatial and temporal constraints of UAVs and tasks, determines the MDC task allocation and tracks the completion of tasks.

        Agents in the air and at sea are mobile. The MDC tasks have the spatial and temporal constraints, requiring UAVs to execute them at specified positions. Therefore, the mobility of UAVs and the spatial and temporal constraints of tasks provide the basis for the MDC task allocation. The MDC task allocation pattern is server allocated tasks (SAT) \cite{27}. The maritime work station collects positions of UAVs and task requirements, and then allocates executable tasks to UAVs through global optimization.

\begin{figure*}
	\centering
	\includegraphics[width=1\linewidth]{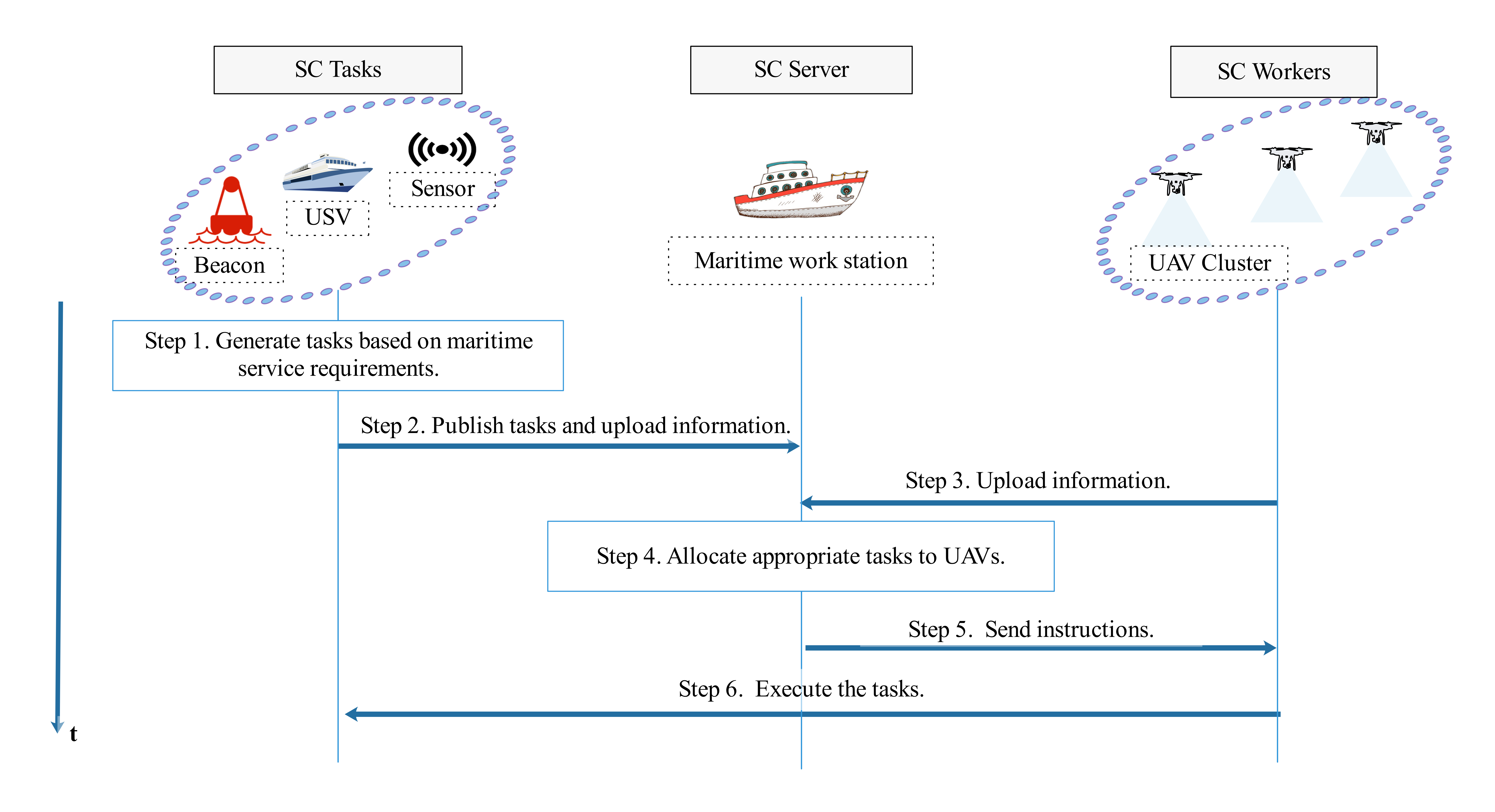}
	\caption{The porposed SC-based MDC task allocation procedure.}\label{MDCFlowChart}
\end{figure*}

        Fig. \ref{MDCFlowChart} shows the SC-based MDC task allocation procedure, which is divided into six steps.
	\begin{enumerate}
		\item	Step 1: The task requesters generate tasks based on maritime service requirements.
		\item	Step 2: The task requesters broadcast tasks to the maritime work station. A task contains data such as its geographical position, the amount of data to be transmitted and the expiration time. The task requesters can upload data to the maritime work station through direct transmission or relay transmission.
		\item	Step 3: UAVs upload data such as their geographical positions, power and energy to the maritime work station. UAVs start from fixed positions and upload data to the maritime work station through direct transmission or relay transmission. 
		\item	Step 4: The maritime work station allocates executable tasks to UAVs. Specifically, the maritime work station analyzes all the collected spatial and temporal data about tasks and UAVs, but does not disclose tasks data to UAVs. Furthermore, the maritime work station evaluates the physical distance and channel conditions between UAVs and the task requesters, calculates the UAV energy consumption and the task waiting time. According to the constraints and actual conditions, the maritime work station allocates executable tasks to UAVs.
		\item	Step 5: The maritime work station sends instructions to UAVs via direct transmission or relay transmission.
		\item	Step 6: UAVs fly to specific positions to execute specific tasks, that is, to collect maritime data. After completing tasks, UAVs need to return to the initial positions to unload the data.
	\end{enumerate}
	
\subsection{Problem Formulation}
	
        Assuming that there are $n$ task requesters issuing commands to the maritime work station to execute tasks, while simultaneously dispatching $m$ UAVs to perform the tasks. The task set to be allocated by the maritime work station is $ \bm{\tau} = [ {{{\tau}_1}, {{\tau}_2},..., {{\tau}_n}} ]$, and the task requester position set is $\bm{\rho}^{\tau}= [{{\rho}^{\tau}_1}, {{\rho}^{\tau}_2},..., {{\rho}^{\tau}_n} ] $. For the requester of the $j$-th $(1 \le j \le n)$ task, its task index is ${{\tau}_j}$, and its position is ${\rho}^{\tau}_j$, denoted as $(x^{\tau}_j,y^{\tau}_j,h^{\tau}_j)$. Since tasks are distributed at sea surface, $h^{\tau}_j$ is $0$. The UAV set is $\bm{\vartheta} = [{{\vartheta}_1}, {{\vartheta}_2},..., {{\vartheta}_m} ]$, the UAV initial position set is $\bm{\rho}^{\vartheta,0}=[{\rho}^{\vartheta,0}_1, {\rho}^{\vartheta,0}_2,..., {\rho}^{\vartheta,0}_m ] $, and the UAV real-time position set is $\bm{\rho}^{\vartheta} = [{\rho}^{\vartheta}_1, {\rho}^{\vartheta}_2,..., {\rho}^{\vartheta}_m ]$. For the $i$-th $(1 \le i \le m)$ UAV, its index is ${{\vartheta}_i}$, its initial position is ${\rho}^{\vartheta,0}_i$, denoted as $(x^{\vartheta,0}_i,y^{\vartheta,0}_i,h^{\vartheta,0}_i)$ and its real-time position is ${\rho}^{\vartheta}_i$, denoted as $(x^{\vartheta}_i,y^{\vartheta}_i,h^{\vartheta}_i)$, where $h^{\vartheta}_i$ is the flying altitude of the UAV. We set the UAV coverage area radius as $r_{{\rm{th}}}$. Therefore, the $j$-th task requester is covered by the $i$-th UAV can be expressed as
	\begin{equation}
		d_{ij}^0 \le r_{{\rm{th}}}, \forall i, \forall j
		\label{dij0th}
	\end{equation}
        where $d_{ij}^0$ is the initial distance between the $j$-th task requester and the $i$-th UAV initial position, which can be expressed as
	\begin{equation}
		d_{ij}^0 = \sqrt {{{\left( {x^{\vartheta,0}_i - x^{\tau}_j} \right)}^2} + {{\left( {y^{\vartheta,0}_i - y^{\tau}_j} \right)}^2}}, \forall i, \forall j
		\label{dij0}
	\end{equation}
	
        The channel conditions between UAVs and task requesters greatly affect the data transmission quality of MDC tasks. and determine the MDC task completion quality. We measure the data transmission quality of MDC tasks by calculating SINR. When the $i$-th UAV executes the $j$-th task, the SINR can be expressed as
	\begin{equation}
		\Phi_{ij} = \frac{{{P_j}\frac{{4\pi }}{{\lambda \sqrt G }}\frac{1}{{d_{ij}^2}}}}{{\left( {I + {\sigma ^2}} \right)}}, \forall i, \forall j
		\label{sinrij}
	\end{equation}
        where ${P_j}$ is the transmit power of the $j$-th task requester, $I$ is the interference when other UAVs collect data, and $\sigma ^2$ is the noise power of the UAV communication channels. $\frac{{4\pi }}{{\lambda \sqrt G }}\frac{1}{{d_{ij}^2}}$ is the path loss, $\frac{{4\pi }}{{\lambda \sqrt G }}$ is the channel power gain at reference distance $d=1 \rm{m}$, $G$ is the antenna orientation coefficient of the $j$-th task requester and $\lambda$ is the wavelength. $d_{ij}$ represents the real-time distance between the $j$-th task requester and the $i$-th UAV real-time position, which can be expressed as
	\begin{equation}
		d_{ij} = \sqrt {{{\left( {x^{\vartheta}_i - x^{\tau}_j} \right)}^2} + {{\left( {y^{\vartheta}_i - y^{\tau}_j} \right)}^2}}, \forall i, \forall j 
		\label{dij}
	\end{equation}
        Further, we set the SINR threshold as $\Phi_{\rm{th}}$ to ensure the data transmission quality of MDC tasks. During the $j$-th task execution, the SINR at the $i$-th UAV is not less than the threshold, which can be expressed as
	\begin{equation}
		\Phi_{ij} \ge \Phi_{\rm{th}}, \forall i, \forall j
		\label{sinrth}
	\end{equation}

        One key issue in the MDC task allocation is that UAVs are energy-constrained. Therefore, the UAV energy consumption model is crucial. The communication energy consumption includes signal processing, circuits, and power amplification. It is worth noting that UAVs also have the propulsion energy consumption to maintain hovering and free movement in the air. Depending on the size and payload of UAVs, the propulsion energy consumption is much greater than the communication energy consumption. In the MDC task allocation, we ignore the communication energy consumption.

        In \cite{zengyong1, zengyong2, zengyong3}, rigorous mathematical derivations have been conducted to obtain the UAV energy consumption model. For a UAV flying horizontally at a speed $v$, the propulsion power can be expressed as
   \begin{equation}
        \begin{aligned}
           P(v) &= P_0(1+3\frac{v^2}{U_{\rm{tip}}}) + P_{\rm{ind}}(1+\frac{v^2}{4v_0^2}-\sqrt{\frac{v^2}{2v_0^2}})^{1/2}\\
           &\quad + \frac{1}{2} A d_0\omega s v^3
           \label{eij}
        \end{aligned}
   \end{equation}
        where $P_0$ and $P_{\rm{ind}}$ are constants, representing the blade profile power and induced power in hover, respectively. They are related to the aircraft weight, air density $\omega$, and rotor disk area $A$. $U_{\rm{tip}}$ represents the blade tip speed, $v_0$ is referred to the average rotor-induced speed in hover. $d_0$ and $s$ are the body drag ratio and rotor solidity, respectively.

        When the $i$-th UAV executes the $j$-th task, the energy consumption can be expressed as
\begin{equation}
E_{ij} = \int P(v_i) \, \rm{d}\mathit{t}_\mathit{i}^{\rm{fly}} + \int \mathit{P}(0) \, \rm{d}\mathit{t}_\mathit{i}^{\rm{hover}}, \quad \forall \mathit{i}, \forall \mathit{j}
\label{eij}
\end{equation}
        where $v_i$ is the $i$-th UAV flight speed. $t_i^{{\rm{fly}}}$ and $t_i^{{\rm{hover}}}$ represent the $i$-th UAV flight time and hover time, respectively \cite{41}. It is worth noting that the UAV energy consumption does not vary significantly at different altitudes, from 50 m to 100 m. In this work, we consider that UAVs perform tasks at specific altitudes \cite{bowen1,bowen2}.

        After collecting data from the specified positions, UAVs are required to return to their initial positions. When the $i$-th UAV returns from the $j$-th task, the UAV energy consumption can be expressed as
	\begin{equation}
		E_{ij}^{0} = \int P(v_i)  {\rm{d}}t_{ij}^{0},\forall i,\forall j
	\end{equation}
        where $t_{ij}^{0}$ is the $i$-th UAV return time, which can be expressed as
	\begin{equation}
		t_{ij}^{0} = \frac{{d_{ij}^{0}}}{{{v_i}}},\forall i,\forall j
	\end{equation}
        We set the UAV energy consumption threshold as $E_{{\rm{th}}}$. 
        Therefore, When the $i$-th UAV executes the $j$-th task, the UAV energy consumption should not exceed the threshold, which can be expressed as
	\begin{equation}
		{E_{ij}} \le E_{{\rm{th}}} = {E^i} - E_{ij}^{0}, \forall i, \forall j
	\end{equation}
        where ${E^i}$ is the $i$-th UAV real-time energy.

        MDC tasks are time-constrained, and they need to be executed in the validity period. The task waiting time refers to the duration between the a task generation and its execution. When the $i$-th UAV executes the $j$-th task, the task waiting time can be expressed as
	\begin{equation}
		{t_{ij}^{\rm{wait}}} = \frac{{{d_{ij}}}}{{{v_i}}}, \forall i, \forall j
		\label{tij}
	\end{equation}
        Furthermore, reducing the task waiting time as much as possible can enhance the MDC task execution efficiency. To ensure the task completion in the validity period, we establish an upper threshold for the task waiting time, denoted as $t_{\rm{th}}^{{\rm{wait}}}$. Specifically, when the $i$-th UAV executes the $j$-th task, the task waiting time should not exceed this threshold, which can be expressed as
	\begin{equation}
		t_{ij}^{{\rm{wait}}} \le t_{{\rm{th}}}^{{\rm{wait}}} = {t^j} - t_j^{{\rm{tran}}},\forall i,\forall j
		\label{tijth}
	\end{equation}
        where ${t^j}$ is the $j$-th task remaining valid time and $t_j^{{\rm{tran}}}$ is the $j$-th task data transmission time, which is generated according to the maritime service requirements.

        When all UAVs and MDC tasks are matched, the $i$-th UAV has $n_i$ tasks to perform. We define the UAV-task match set as ${\bm{\Xi}} = [ {\bm{\Xi}}^1, {\bm{\Xi}}^2,..., {\bm{\Xi}}^m ]$, and the task completion time set as ${\bm{T}^{\vartheta}} = [ t^{\vartheta}_1,t^{\vartheta}_2,...,t^{\vartheta}_m]$. For the $i$-th UAV, its match set with tasks is $\bm{{\Xi}^i} =[  {\tau}_1^i,{\tau}_2^i,..., {\tau}_{{n^i}}^i ]$. $ {\tau}_k^i $ ( $ 0 \le k \le {n^i} $) represents the $k$-th task that the UAV need to execute. The task completion time for the $i$-th UAV can be expressed as
	\begin{equation}
		t_i^{\vartheta} = \sum\limits_{j = 1}^{n_i} {\left( {\frac{{{d_{ij}}}}{{{v_i}}} + t_j^{{\rm{tran}}}} \right)} + t_{i{n_i}}^0, \forall i
		\label{ti} 
	\end{equation}
        Furthermore, the energy consumption for the $i$-th UAV can be expressed as
	\begin{equation}
		E_i = \sum\limits_{j = 1}^{n_i} {\left( E_{ij} \right)} + E_{i{n_i}}^0, \forall i
	\end{equation}     

        The maritime work station analyzes the spatial and temporal constraints of all UAVs and MDC tasks, then assists in UAV-task match. Based on the matching results, UAVs proceed to execute the corresponding tasks. Under the condition that tasks are executed in the validity period, the maritime work station shortens the task completion time. The corresponding optimization problem can be formulated as follows
	\begin{align}
		\mathop {{\rm{minimize}}}\limits_{\bm{\Xi}}  
        & \frac{1}{m}\sum\limits_{i = 1}^m {t_i^{\vartheta}}\label{problem} \\
		{\rm{Subject\ to}}\ 
        & \ d_{ij}^0 \le r_{{\rm{th}}}, \forall i, \forall j \label{prodij}\tag{\ref{problem}{a}} \\
		&\Phi_{ij} \ge \Phi_{\rm{th}}, \forall i, \forall j \label{prosinr}\tag{\ref{problem}{b}}\\ 
		&{E_{ij}} \le E_{{\rm{th}}}, \forall i, \forall j \label{proeij}\tag{\ref{problem}{c}}\\
		&t_{ij}^{{\rm{wait}}} \le t_{{\rm{th}}}^{{\rm{wait}}}, \forall i, \forall j \label{protij}\tag{\ref{problem}{d}}\\
		&{\tau}_k^i \in \tau, \forall k, \forall i  \label{proski}\tag{\ref{problem}{e}}\\
		&0 \le \sum\limits_{i = 1}^m {{n_i} \le n} \label{prosum}\tag{\ref{problem}{f}}\\
		&i \in { 1,2,...,m} \label{i}\tag{\ref{problem}{g}}\\
		&j \in { 1,2,...,n} \label{j}\tag{\ref{problem}{h}}\\
		&k \in {1,2,...,n^i} \label{k}\tag{\ref{problem}{i}}
	\end{align}    
        The optimization objective of (\ref{problem}) is to minimize the the average task completion time for all UAVs by the UAV-task match. Constraint (\ref{prodij}) ensures the tasks executed by a UAV are in its coverage area. Constraint (\ref{prosinr}) ensures the data transmission quality of MDC tasks. Constraint (\ref{proeij}) guarantees that UAVs have enough energy to perform their tasks and return to their initial positions. Constraint (\ref{protij}) ensures that the task waiting time does not exceed a predefined threshold. Constraint (\ref{proski}) requires all tasks executed by UAVs must be present in the task set, while constraint (\ref{prosum}) ensures the total number of tasks executed by all UAVs does not exceed the number of task requesters. Constraint (\ref{i}), (\ref{j}), (\ref{k}) respectively define the possible value ranges for variables $i$, $j$, and $k$.

\section{Proposed Solution}

        In this section, we propose the SC-MDC-TA algorithm to address the formulated optimization problem in (\ref{problem}), which achieves the MDC task allocation in different maritime service scenarios. Specifically, considering the UAV capability, we assess and control the MDC task completion quality with the quality estimation algorithm. Meanwhile, considering the task waiting time, we adopt the reverse auction algorithm to shorten the MDC task completion time as much as possible on the condition that tasks are executed in the validity period.

        We set the UAV-task match set at time slot $t$ as $\bm{M}^{t} = [m_{ij}]_{m \times n}$, which is time-varying. If the $i$-th UAV is matched with the $j$-th task, $m_{ij}$ is set to 1; otherwise, it is set to 0. According to equation \ref{dij0}, we establish the initial distance set between UAVs and tasks as ${{\bm{D}}^0}=[d^0_{ij}]_{m\times n}$. Following equation \ref{dij}, we define the real-time distance set between UAVs and tasks as ${\bm{D}}=[d_{ij}]_{m\times n}$. Moreover, we set the UAV state set as ${\bm{\chi}^{\vartheta}=[{\chi}^{\vartheta}_{i}]_{1\times m}}$. If the $i$-th UAV is idle, ${\chi}^{\vartheta}_{i}$ is set to 0; otherwise, it is set to 1. Similarly, we define the task state set as ${\bm{\chi}^{\tau}=[{\chi}^{\tau}_{j}]_{1\times N}}$. If the $j$-th task is unexecuted, ${\chi}^{\tau}_{j}$ is set to 0; otherwise, it is set to 1.

\subsection{SC-MDC-TA Algorithm}

\begin{algorithm}[t]
	\caption{SC-MDC-TA algorithm}	
	\begin{algorithmic}[1] 
	\State {\bf Input:}  $\bm{\rho}^{\vartheta, 0}$, $\bm{\rho}^{\tau}$.
	\State {\bf Output:}  $\bm{\Xi}$, $\bm{T}$, $\bm{E}$.
	\Statex 
	\State Initialize $t$, $\bm{\chi}^{\vartheta}$, $\bm{\chi}^{\tau}$, $\bm{\Xi}$, $\bm{T}$, $\bm{E}$.

	\State Set $\bm{T}^{{\rm{tran}}}$, $\bm{H}^{\vartheta}$, $\bm{V}$, $\bm{E}^{\vartheta}$, $r_{{\rm{th}}}$, $r_{{\rm{com}}}$, $\Phi_{\rm{th}}$, $E_{{\rm{th}}}$, $t_{{\rm{th}}}^{{\rm{wait}}}$.

	\State Calculate the initial distance set $\bm{D}^0$ according to equation (\ref{dij0}).			
	\For{$t$}	
	\State Select idle UAVs and unexecuted tasks according to $\bm{\chi}^{\vartheta}$ and $\bm{\chi}^{\tau}$.
	\State Update $\bm{D}$.
	\State Initialize  ${\bm{M}}^{t}$.
	\While{not all idle UAVs are matched to a task}
			\State Update ${\bm{M}}^{t}$ according to \textbf{Algorithm \ref{algorithm1}}.
			\State Update ${\bm{M}}^{t}$ according to \textbf{Algorithm \ref{algorithm2}}.
			\If{${\bm{M}}^{t}[a,:]$ has only one element equal to 1}
				\State Find column $a'$ of element 1.
				\State The $a$-th UAV executes the $a'$-th task.
			\ElsIf{${\bm{M}}^{t}[b,:]$ has multiple elements equal to 1}
				\State Find the column $b'$ of min$\bm{D}[\mathit{b},:]$.
				\State The $b$-th UAV executes the $b'$-th task.
			\EndIf
   
	\EndWhile
	\State Update $\bm{\chi}^{\vartheta}$, $\bm{\chi}^{\tau}$, $\bm{\Xi}$, $\bm{T}$, $\bm{E}$, $\bm{E}^{\vartheta}$, $\bm{\rho}^{\vartheta}$.
 
	\If{all elements in $\bm{\chi}^{\tau}$ are 1} 
	\State Update $\bm{T}$, $\bm{E}$ based on the return distance.
	\State break
          \Else
	\State $t=t+1$.
	\EndIf
	\EndFor
	\end{algorithmic}
	\label{algorithm}
\end{algorithm}

        The proposed SC-MDC-TA algorithm is shown in \textbf{Algorithm \ref{algorithm}}. The inputs include the UAV initial position set $\bm{\rho}^{\vartheta, 0}$ and the task requester position set $\bm{\rho}^{\tau}$. The outputs are the UAV-task match set $\bm{\Xi}$, the task execution time set $\bm{T}=[t^{\vartheta}_i]_{1\times m}$, and the UAV energy consumption set $\bm{E}=[E_i]_{1\times m}$.

        Specifically, we initialize the time slot $t=0$ and set the elements of $\bm{\chi}^{\vartheta}$, $\bm{\chi}^{\tau}$, $\bm{\Xi}$, $\bm{T}$, and  $\bm{E}$ to zero. Additionally, we set the task transmission time set $\bm{T}^{{\rm{tran}}}=[t^{\rm{tran}}_i]_{1\times m}$, the UAV flight height set $\bm{H}^{\vartheta}=[h^{\vartheta}_i]_{1\times m}$, the UAV speed set $\bm{V}=[v_i]_{1\times m}$, the UAV initial energy set $\bm{E}^{\vartheta}=[E^i]_{1\times m}$, the UAV coverage area radius $r_{{\rm{th}}}$, the UAV communication radius $r_{{\rm{com}}}$, the SINR threshold $\Phi_{\rm{th}}$, the energy threshold $E_{{\rm{th}}}$, and the task waiting time threshold $t_{{\rm{th}}}^{{\rm{wait}}}$. The initial distance set between UAVs and tasks $\bm{D}^0$ is calculated according to equation (\ref{dij0}).
     
        In each time slot, the current idle UAVs and unexecuted tasks are selected according to $\bm{\chi}^{\vartheta}$ and $\bm{\chi}^{\tau}$, and the real-time distance set ${\bm{D}}$ is calculated. The UAV-task match set at time slot $t$ $\bm{M}^t$ is initialized, and the following steps are repeated until all currently idle UAVs are matched to a task for execution.
     
        To ensure the task completion quality, \textbf{Algorithm \ref{algorithm1}} selects MDC tasks that can be executed by UAVs based on $r_{{\rm{th}}}$, $r_{{\rm{com}}}$, $\Phi_{ij}$, and ${E_{\rm{ij}}}$. When multiple UAVs select the same MDC task, \textbf{Algorithm \ref{algorithm2}} preferentially allocates the task to the UAV with the shortest task waiting time. Each UAV can handle only one task at a time. If UAV $a$ selects only one task, it will execute that task. If UAV $b$ selects multiple tasks, it will execute the nearest one. Other tasks are allocated to other idle UAVs at this time or the next time.

        Based on the allocation results, we update $\bm{\chi}^{\vartheta}$, $\bm{\chi}^{\tau}$, $\bm{\Xi}$, $\bm{T}$, $\bm{E}$, $\bm{E}^{\vartheta}$, $\bm{\rho}^{\vartheta}$. If there are no executable tasks in the coverage area, the UAV returns to its initial point. If all tasks are allocated, update $\bm{T}$ and $\bm{E}$ based on the UAV return distance, and the algorithm concludes. Otherwise, proceed to the next time slot.

\begin{algorithm}[t]
    \caption{Quality estimation algorithm}
    \begin{algorithmic}[1]
    \State {\bf Input:} ${\bm{M}}^{t}$, $\bm{D}^0$, $\bm{D}$, $r_{{\rm{th}}}$, $r_{{\rm{com}}}$, $\Phi_{\rm{th}}$, $E_{{\rm{th}}}$.
    \State {\bf Output:}  ${\bm{M}}^{t}$.  
    \Statex 
        \For{$i$ \textbf{in} idle UAVs}
            \For{$j$ \textbf{in} unexecuted tasks}
                \State Calculate $\Phi_{ij}$ and ${E_{ij}}$ based on $\bm{D}$.
                \If{$d_{ij}^0\le r_{{\rm{th}}}$ \& $d_{ij}\le r_{{\rm{com}}}$ \& $\Phi_{ij} \ge \Phi_{\rm{th}}$ 
                \Statex \quad \quad \quad \quad \quad \quad \& ${E_{ij}} \le E_{{\rm{th}}}$}
                    \State ${m_{ij}} = 1$.
                \Else
                    \State ${m_{ij}} = 0$.
                \EndIf
                \If{all elements in ${\bm{M}}^{t}[i,:]$ are 0}
                \State Find column $d$ of min$\bm{D}[i,:]$.
                \State ${m_{id}} = 1$.
                \EndIf
            \EndFor
        \EndFor        
    \end{algorithmic}
    \label{algorithm1}
\end{algorithm}
	
	\begin{table}[t]
		\centering
		\caption{The MDC task allocation based on the quality estimation algorithm.}
		\begin{tabular}{cc}
			\hline
			\textbf{Index of idle UAVs}	& \textbf{Index of executable tasks}\\
			\hline
			UAV1		& Task1, Task2, Task4……		\\
			UAV2		& Task2, Task4, Task5……		\\
			UAV4		& Task4, Task5, Task6……		\\
			\hline
			\label{matchquality}
		\end{tabular}
	\end{table}
 
\subsection{Quality Estimation Algorithm}
 
        In the MDC task allocation, serving as data collection platforms, UAVs gather data generated by various agents on the sea surface. UAVs have no certain requirements on the task sequence. Nonetheless, tasks published by the task requesters have time constraints, as they need to return data as soon as possible in the validity period. Due to the lack of objective consistency between UAVs and task requesters, we design the quality estimation to ensure the MDC task completion quality.

        The quality estimation algorithm is shown in \textbf{Algorithm \ref{algorithm1}}. For the current idle UAVs and unexecuted tasks, the maritime work station calculates $\Phi_{ij}$ and ${E_{ij}}$. When the $i$-th UAV executes the $j$-th task, we make the following judgments: 
\begin{enumerate}
    \item whether the initial distance ${d_{ij}^0}$ is not greater than the UAV coverage area radius $r_{{\rm{th}}}$;
    \item whether the real-time distance ${d_{ij}}$ is not greater than the UAV communication radius $r_{{\rm{com}}}$; 
    \item whether $\Phi_{ij}$ is not less than the SINR threshold $\Phi_{\rm{th}}$; 
    \item whether ${E_{ij}}$ is not greater than the energy threshold $E_{{\rm{th}}}$.
\end{enumerate}   
        If the UAV and task meet these conditions, we consider that the UAV can guarantee the task completion quality. Thus, the UAV is matched with this task. If there is no task that meets the conditions for the UAV at this time, the UAV will be matched with the nearest UAV.

        Through the quality estimation algorithm, the maritime work station allocates executable tasks to current idle UAVs. It is possible that a UAV select multiple tasks. The MDC task allocation examples based on this algorithm are shown in Table \ref{matchquality}. The first column indicates the index of idle UAVs at this time slot, and the second column indicates the index of executable tasks.

\subsection{Reverse Auction Algorithm}

        Appropriate rewards are conducive to the optimal task allocation and can facilitate the MDC task effective completion. Poor rewards will reduce UAVs' enthusiasm and task completion efficiency. However, high rewards not only hurt the task completion amount, but also reduce the overall efficiency of the maritime work station. We design the reverse auction algorithm to further improve the MDC task allocation, thereby boosting UAVs' enthusiasm to execute tasks and reducing the task waiting time. In the reverse auction, tasks are central while the task waiting time is the auction item. The maritime work station acts as the auctioneer and UAVs act as buyers that submit bids to it, with the submitted bids are the task waiting time. When multiple UAVs select a task, these UAVs initiate bidding requests to the maritime work station.

        The reverse auction algorithm is shown in \textbf{Algorithm \ref{algorithm2}}. For the current idle UAVs and their executable tasks, the maritime work station calculates the waiting time $t_{ij}^{{\rm{wait}}}$. When the $i$-th UAV executes the $j$-th task, if the task waiting time exceeds the threshold $t_{{\rm{th}}}^{{\rm{wait}}}$, the matching between them is canceled. Additionally, if a task is matched to multiple UAVs, the maritime work station selects the UAV with the shortest task waiting time as the optimal match for the task.

        Through the reverse auction algorithm, the maritime work station allocates MDC tasks to the optimal UAV for execution. This ensures that MDC tasks are not only executed in their validity period but also completed as quickly as possible. Building upon the basis of Table \ref{matchquality}, the MDC task matching example based on the reverse auction algorithm is shown in Table \ref{matchreverse}. The first column represents the index of unexecuted tasks, the second column represents the index of idle UAVs, and the third column represents the index of the optimal selected UAV.

\subsection{Algorithm Complexity Analysis}

        In this subsection, we analyze the computational complexity of the proposed SC-MDC-TA algorithm. The algorithm mainly consists of an outer loop and an inner loop. The outer loop serves as the primary loop of the entire algorithm and is controlled by the number of iterations, denoted as $t$. In each iteration of the outer loop, two computations and one inner loop are performed. Firstly, the time complexity of updating the distance between tasks and real-time positions of UAVs is $O(m \cdot n)$. Secondly, the time complexity of updating various parameters of UAVs based on the results of the inner loop is $O(m)$. The inner loop is responsible for allocating tasks to current idle UAVs, with a maximum iteration count of $m$ in the worst case. In each iteration of the inner loop, two main computations are performed: the quality estimation algorithm and the reverse auction algorithm. The time complexity of the quality estimation algorithm, which calculates ${\Phi}_{ij}$ and ${E_{ij}}$, is $O(A \cdot B)$, where $A$ represents the number of idle UAVs and $B$ represents the number of unexecuted tasks. The reverse auction algorithm computes $t_{ij}^{{\rm{wait}}}$ and has a time complexity of $O(A \cdot B)$. Since $A \le m$ and $B \le n$, the time complexity of the inner loop is $O(m \cdot (m \cdot n))$. In conclusion, the overall time complexity of the entire algorithm is $O(t \cdot m \cdot m \cdot n)$. If $t$, $m$, and $n$ increase, the execution time of the algorithm will correspondingly increase.

\begin{algorithm}[t]
    \caption{Reverse auction algorithm}
    \begin{algorithmic}[1] 
    \State {\bf Input:}\  ${\bm{M}}^{t}$, $\bm{D}$, $t_{{\rm{th}}}^{{\rm{wait}}}$.
    \State {\bf Output:}\  ${\bm{M}}^{t}$.
    \Statex
        \For{$j$ \textbf{in} unexecuted tasks}
            \For{$i$ \textbf{in} idle UAVs}
                \State Calculate $t_{ij}^{{\rm{wait}}}$ based on $\bm{D}$.
                \If{$t_{ij}^{{\rm{wait}}} > t_{{\rm{th}}}^{{\rm{wait}}}$}
                    \State ${\bm{M}}^{t}[i,j]=0$
                \EndIf
            \EndFor
            \If{${\bm{M}}^{t}$ has multiple elements equal to 1 in the $j$-th column}
                \State ${\bm{M}}^{t}[:,j]=0$
                \State Set $m_{ij}=1$ for the $i$-th row with the minimum $t_{ij}^{{\rm{wait}}}$
            \EndIf
        \EndFor
    \end{algorithmic}
    \label{algorithm2}
\end{algorithm}

	\begin{table}[t]
		\centering
		\caption{The MDC task allocation based on the reverse auction algorithm.}
		\begin{tabular}{p{2cm} p{3cm} p{2.5cm}}
			\hline
			\textbf{Index of unexecuted tasks}	& \textbf{Index of idle UAVs}&\textbf{Index of the optimal selected UAV}	\\
			\hline
			Task2		& UAV1, UAV2…	& UAV3	\\
				Task5		& UAV1, UAV4…	& UAV1	\\
				Task7		& UAV2, UAV4…	& UAV4	\\
			\hline
			\label{matchreverse}
		\end{tabular}
	\end{table}
 
\section{Simulation Results and Discussion}

        In this section, simulation parameters are specified. Numerical results are demonstrated to validate the effectiveness of the proposed SC-MDC-TA algorithm.
	
\subsection{Parameter Settings}

	\begin{figure}[b!]
		\centering
		\includegraphics[width=0.9\linewidth]{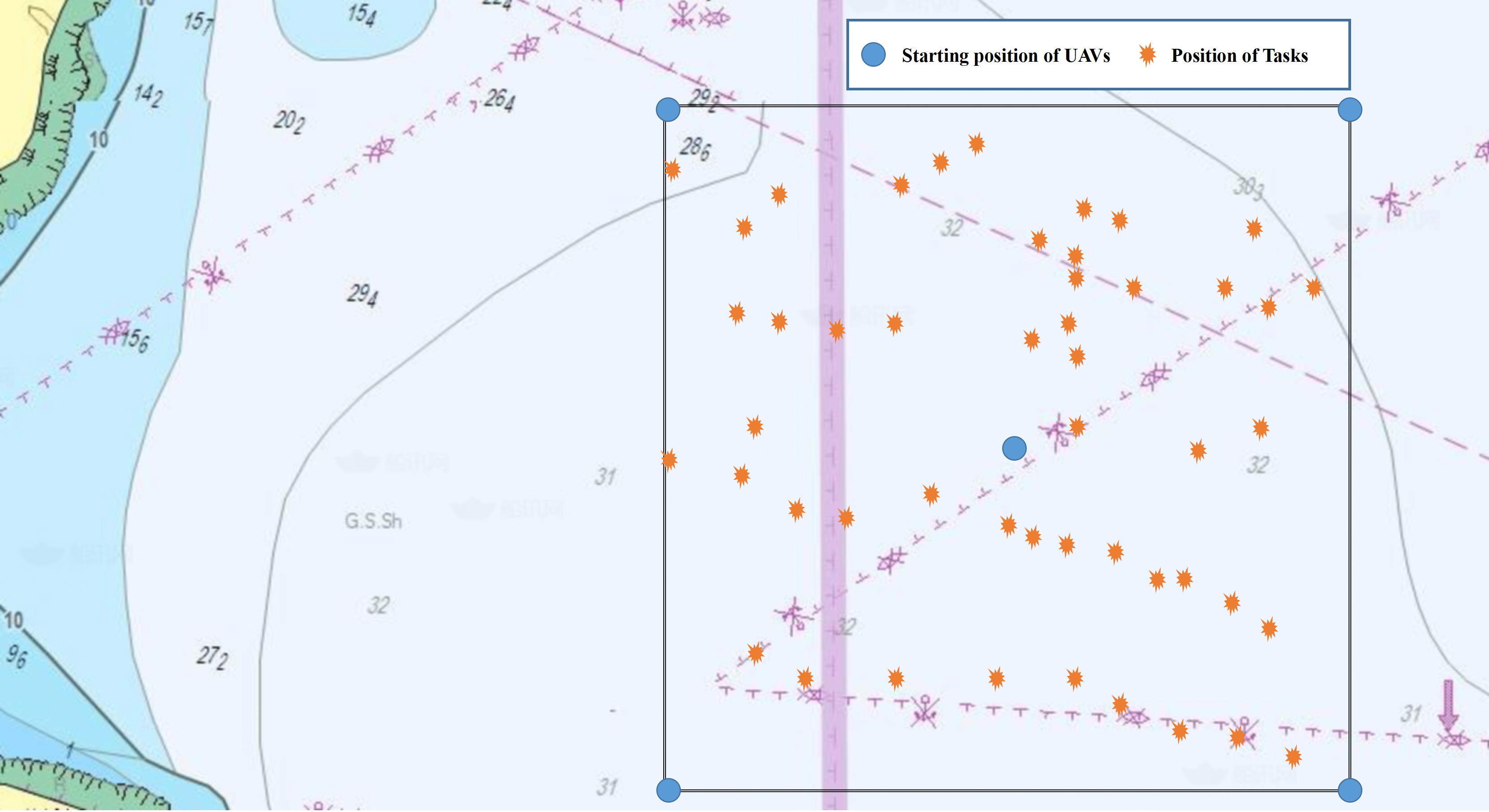}
		\caption{Task positions and UAV positions in the electronic navigational charts.}\label{SimulationScene}
	\end{figure} 
 
        We establish a simulation scenario with an area of $1.5 \  \rm{km} \times 1.5 \  \rm{km}$, which is selected from part of the offshore sea area according to the electronic navigational charts. The selected maritime area, task positions, and UAV positions are shown in the electronic navigation chart as depicted in Fig. \ref{SimulationScene}. The tasks are represented by yellow symbols. The data transmission time $t_j^{{\rm{tran}}}$ for each task is generated based on the data volume in the scenario. Meanwhile, UAVs take off and returns from the positions, executing tasks at a constant flight altitude and speed. Then, we set the task number $n$ to $50$, the UAV number $m$ to $4$, the UAV flight height ${h^{\vartheta}_i}$ to $ 20\ {\rm{m}}$, the UAV speed ${v_i}$ to $ 25\ {\rm{m}}/{\rm{s}}$ \cite{11}. We set the UAV flight power $P_i^{{\rm{fly}}}$ to $ 87\ {\rm{W}}$ and UAV hovering power $P_i^{{\rm{hover}}} $ to $ 60\ {\rm{W}}$ \cite{10}, respectively. Compared with the UAV flight power and hovering power, the UAV communication power is set to $100\ {\rm{mW}}$, which is negligible. We employ the closest distance MDC task allocation algorithm (CD-MDC-TA) as the benchmark. CD-MDC-TA aims to allocate the closest task to each currently idle UAV, thereby achieving the shortest response time. This is achieved by prioritizing the fulfillment of task requesters' needs with the closest available UAVs.

\subsection{Performance Comparison Under Different Hover Time}

  \begin{figure*}
	\centering
	\subfloat[No hover]{\includegraphics[width=0.33\textwidth]{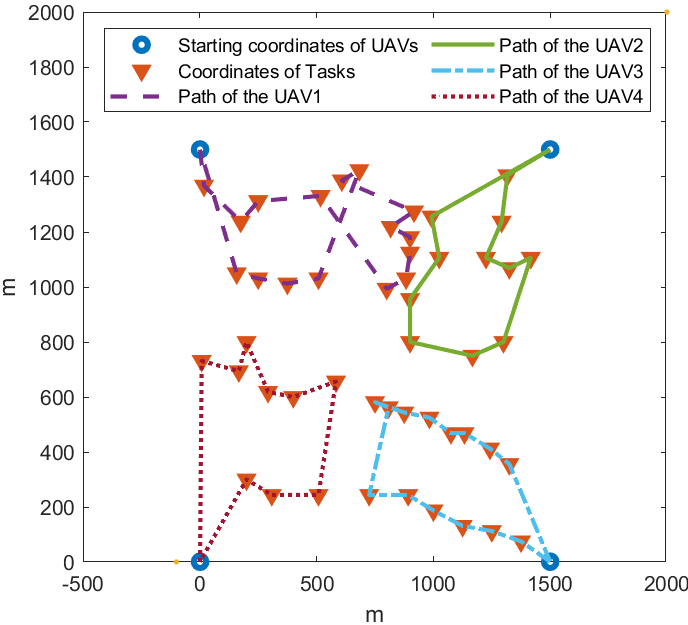}\label{Timesallocation1}} 	
	\subfloat[0-5s]{\includegraphics[width=0.33\textwidth]{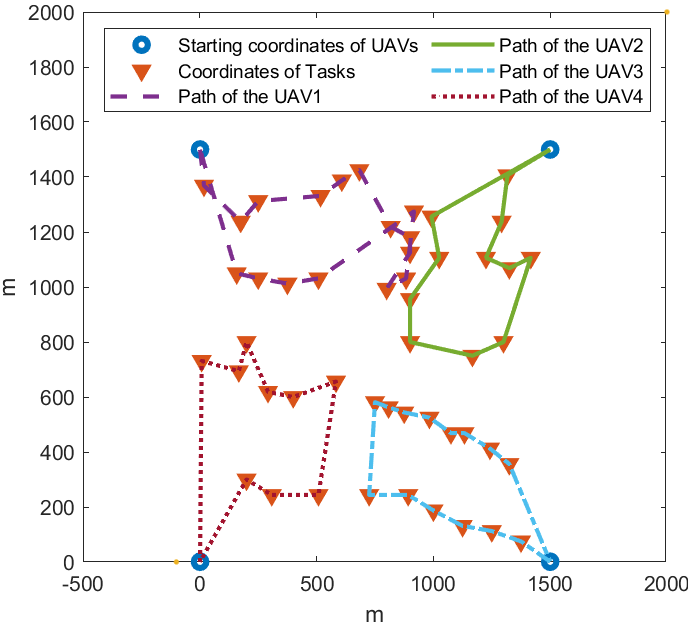}\label{Timesallocation2}}
	\subfloat[10-30s]{\includegraphics[width=0.33\textwidth]{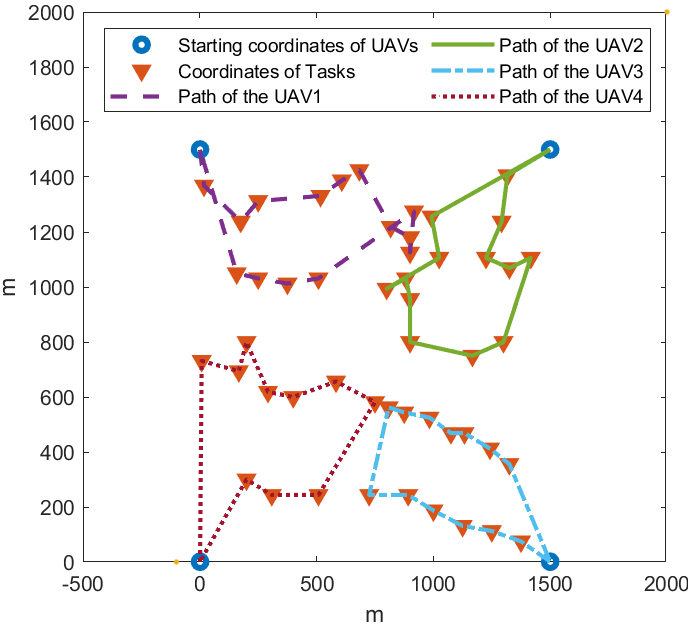}\label{Timesallocation3}} 	
	\caption{The MDC task allocation based on the proposed SC-MDC-TA algorithm under different hover time.}\label{Timesallocation}
\end{figure*}
  
  \begin{figure}
	\centering
	\subfloat[No hover]{\includegraphics[width=0.15\textwidth]{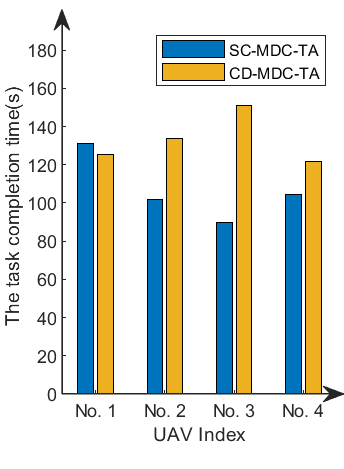}\label{Timetime1}} 	
	\subfloat[0-5s]{\includegraphics[width=0.15\textwidth]{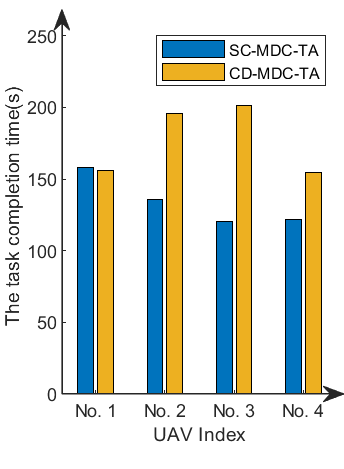}\label{Timetime2}}
	\subfloat[10-30s]{\includegraphics[width=0.15\textwidth]{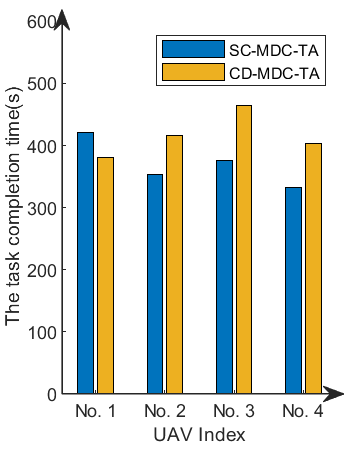}\label{Timetime3}} 	
	\caption{The comparison of task completion time under different hover time.}\label{Timetime}
\end{figure}

  \begin{figure}
	\centering
	\subfloat[No hover]{\includegraphics[width=0.15\textwidth]{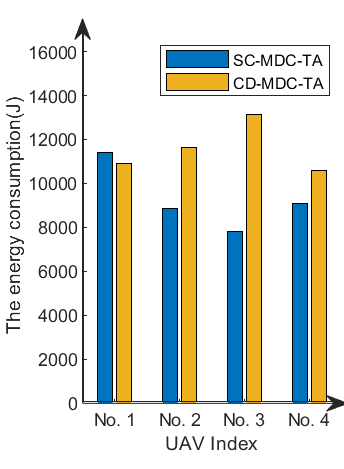}\label{Timeenergy1}} 	
	\subfloat[0-5s]{\includegraphics[width=0.15\textwidth]{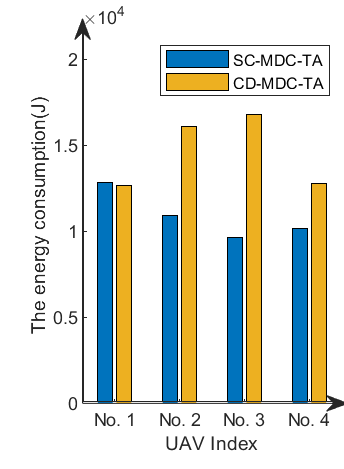}\label{Timeenergy2}}
	\subfloat[10-30s]{\includegraphics[width=0.15\textwidth]{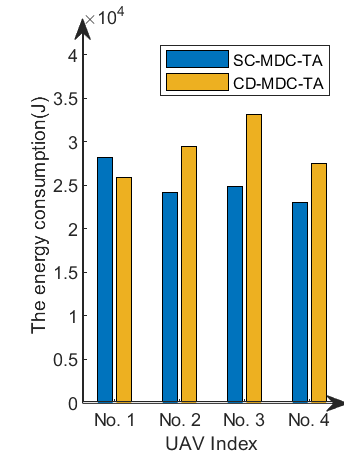}\label{Timeenergy3}} 	
	\caption{The comparison of UAV energy consumption under different hover time.}\label{Timeenergy}
\end{figure}

  \begin{figure}
	\centering
	\subfloat[Task completion time]{\includegraphics[width=0.23\textwidth]{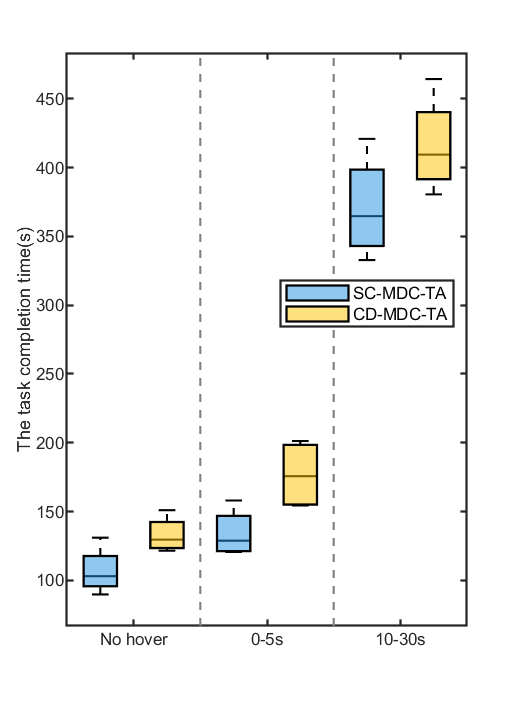}\label{Timecompare1e}} 	
	\subfloat[UAV energy consumption]{\includegraphics[width=0.23\textwidth]{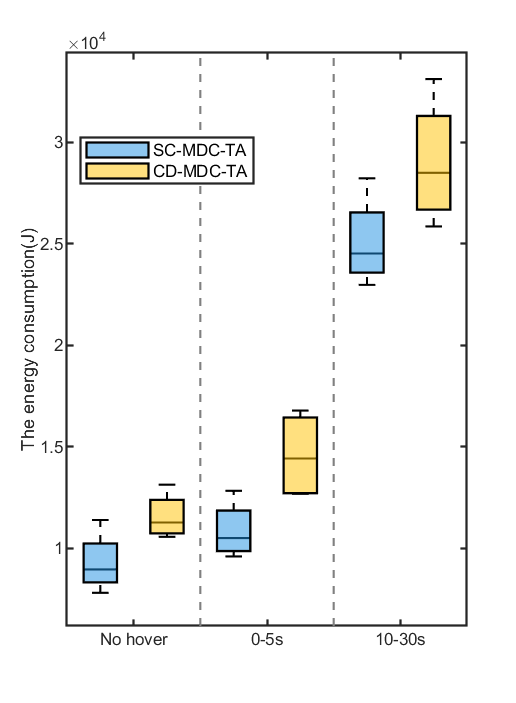}\label{Timecompare2}}	
	\caption{Performance boxplot comparison under different hover time.}\label{Timecompare}
\end{figure}

The hover time of UAVs plays a pivotal role in task execution strategies. When the hover time is set to 0, UAVs operate without pause, prioritizing swift execution speeds and uninterrupted energy efficiency. Adjusting hover times to the 0-5s range allows for momentary stops at designated points, facilitating more accurate data gathering or swift status evaluations. Extending this hover period to 10-30s enables UAVs to engage in extensive data interaction within target zones, particularly suited for scenarios necessitating meticulous monitoring or intricate maneuvers. Fig. \ref{Timesallocation} elucidates how variations in hover times impact the allocation of UAVs for MDC tasks. The proposed SC-MDC-TA algorithm showcases its adaptability and ingenuity by dynamically refining flight path planning in real-time, thereby significantly enhancing the temporal efficiency of task execution. This underscores the broad applicability of the SC-MDC-TA algorithm, which tailors its approach according to the precise hovering necessities of individual tasks, thereby optimizing the overall effectiveness and responsiveness of maritime operations.

Fig. \ref{Timetime} and Fig. \ref{Timeenergy} visually compare the task completion time and energy consumption of the proposed SC-MDC-TA algorithm with the CD-MDC-TA algorithm under different hover time conditions. The proposed SC-MDC-TA algorithm consistently shows superiority, whether there is no hover, brief hover, or longer hover periods. Specifically, in hover conditions of no hover, 0-5s, and 10-30s, the proposed SC-MDC-TA algorithm achieves reductions in task completion time of 19.7\%, 24.2\%, and 10.9\%, respectively, compared to the benchmark. Meanwhile, energy consumption is reduced by 19.9\%, 25.5\%, and 13.6\%, respectively. This indicates that the proposed SC-MDC-TA algorithm not only responds promptly to task demands but also effectively manages the energy budget of UAVs, ensuring the efficiency and cost-effectiveness of task execution.

Fig. \ref{Timecompare} clearly presents the performance distribution differences of the two algorithms in terms of task completion time and UAV energy consumption through box plots. The proposed SC-MDC-TA algorithm shows a more compact data distribution, indicating that it offers a more stable and consistent performance when dealing with tasks with different hover time. The low variance means the algorithm has stronger control over task execution, avoiding significant performance fluctuations due to hover time variations, thus ensuring the reliability of maritime services.

The superior performance of the proposed SC-MDC-TA algorithm is due to its comprehensive consideration of task quality, UAV capabilities, and task timeliness. By introducing a reverse auction mechanism, the algorithm can intelligently match the most suitable UAV to perform specific tasks, effectively reducing task waiting time while ensuring tasks are completed within their validity periods. Additionally, the proposed SC-MDC-TA algorithm optimizes the UAV coverage area, avoiding unnecessary long-distance flights and saving energy. This integrated optimization strategy enables the proposed SC-MDC-TA algorithm to meet the diverse needs of maritime services while achieving the dual improvements in task execution efficiency and energy utilization efficiency, opening up new possibilities for future UAV applications in the maritime domain.

  \begin{figure*}
	\centering
	\subfloat[Center]{\includegraphics[width=0.33\textwidth]{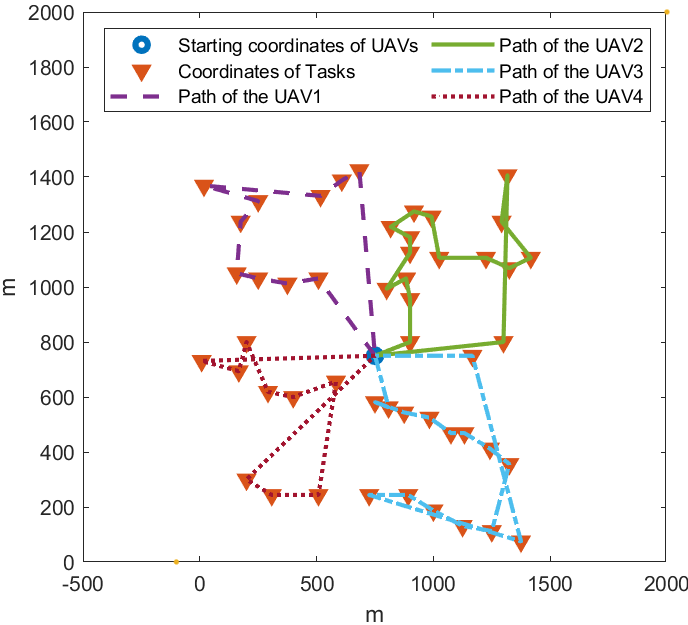}\label{211-1-A}} 	
	\subfloat[Corners]{\includegraphics[width=0.33\textwidth]{figures/212-1-A}\label{212-1-A}}
	\subfloat[Edge]{\includegraphics[width=0.33\textwidth]{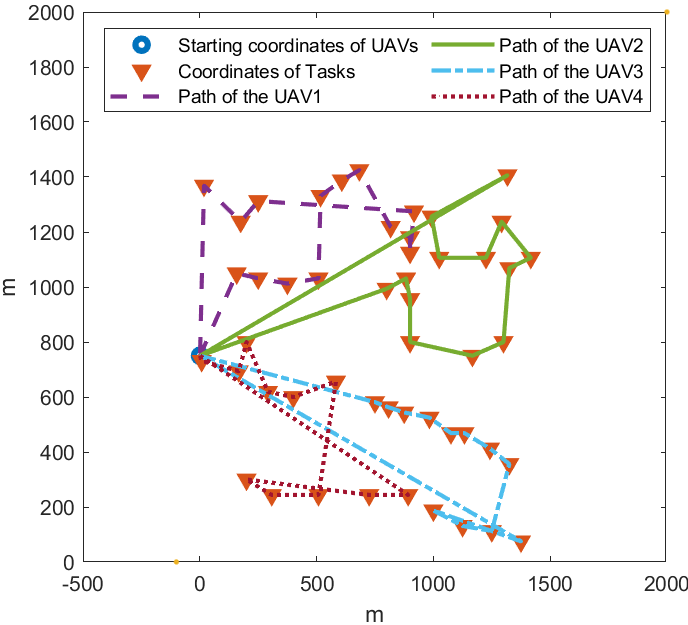}\label{213-1-A}} 	
	\caption{The MDC task allocation based on the proposed SC-MDC-TA algorithm from different take-off positions.}\label{locationsallocation}
\end{figure*}

\begin{figure}
	\centering
	\subfloat[Center]{\includegraphics[width=0.15\textwidth]{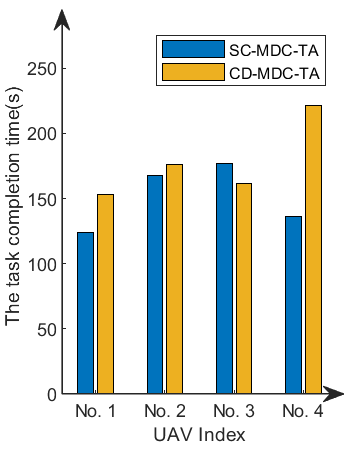}\label{211-2-time}} 	
	\subfloat[Corners]{\includegraphics[width=0.15\textwidth]{figures/212-2-time}\label{212-2-time}}
	\subfloat[Edge]{\includegraphics[width=0.15\textwidth]{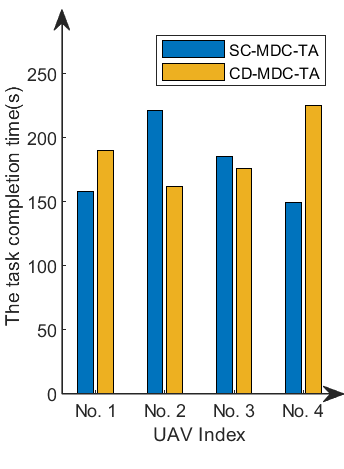}\label{213-2-time}} 	
	\caption{The comparison of task completion time from different take-off positions.}\label{locationtime}
\end{figure}

\begin{figure}
	\centering
	\subfloat[Center]{\includegraphics[width=0.15\textwidth]{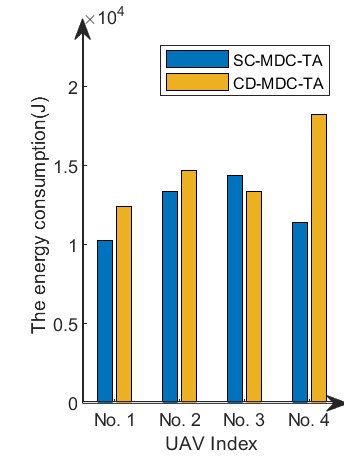}\label{211-3-energy}} 	
	\subfloat[Corners]{\includegraphics[width=0.15\textwidth]{figures/212-3-energy}\label{212-3-energy}}
	\subfloat[Edge]{\includegraphics[width=0.15\textwidth]{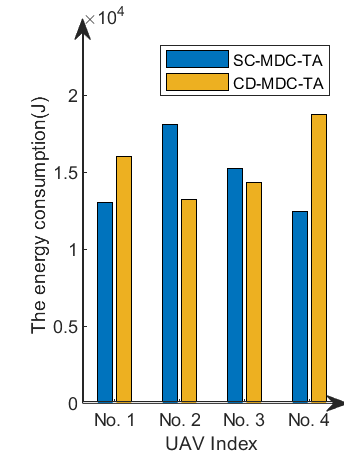}\label{213-3-energy}} 	
	\caption{The comparison of task completion time from different take-off positions.}\label{locationenergy}
\end{figure}

\begin{figure}
	\centering
	\subfloat[Task completion time]{\includegraphics[width=0.23\textwidth]{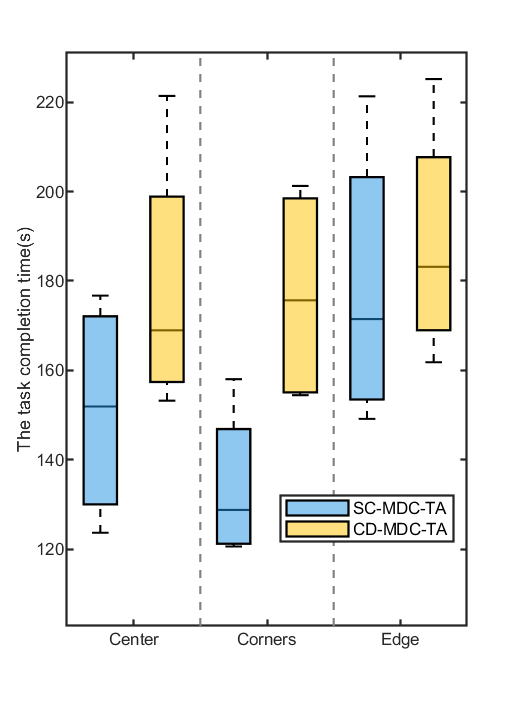}\label{boxtime-time-time}} 	
	\subfloat[UAV energy consumption]{\includegraphics[width=0.23\textwidth]{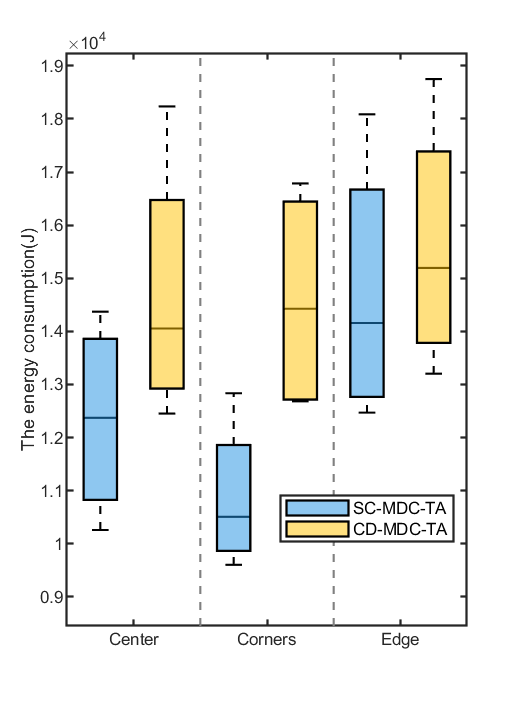}\label{boxtime-time-energy}}	
	\caption{Performance boxplot comparison from different take-off positions.}\label{locationcompare}
\end{figure}

\begin{figure*}
	\centering
	\subfloat[Random]{\includegraphics[width=0.33\textwidth]{figures/212-1-A}\label{211-1-A}} 	
	\subfloat[Uniform]{\includegraphics[width=0.33\textwidth]{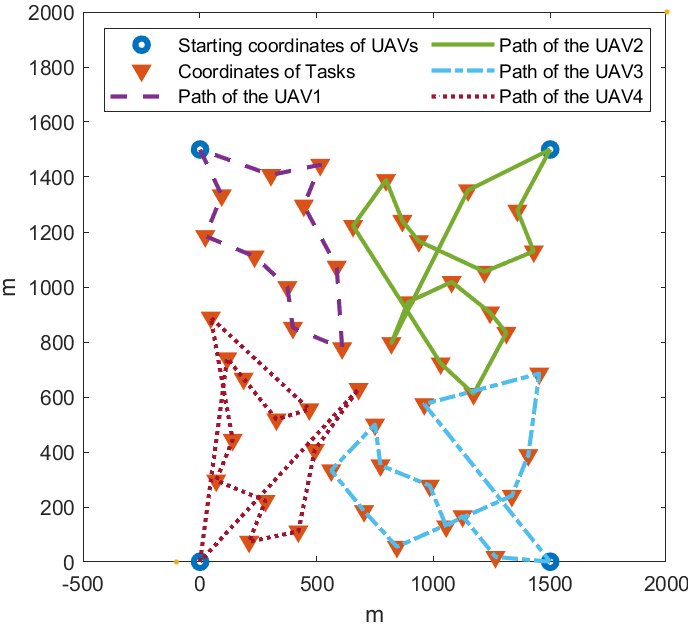}\label{212-1-A}}
	\subfloat[Clustered]{\includegraphics[width=0.33\textwidth]{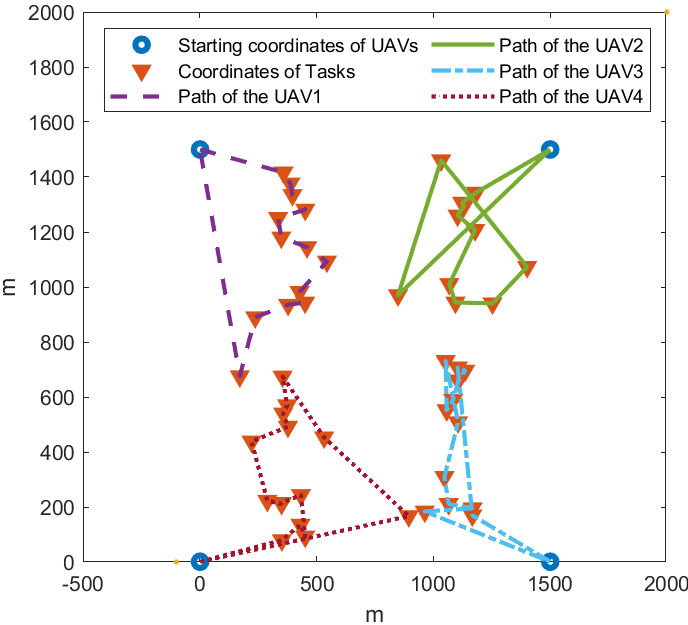}\label{213-1-A}} 	
	\caption{The MDC task allocation based on the proposed SC-MDC-TA algorithm under different task distributions}\label{tasksallocation}
\end{figure*}

\begin{figure}
	\centering
	\subfloat[Random]{\includegraphics[width=0.15\textwidth]{figures/212-2-time}\label{211-2-time}} 	
	\subfloat[Uniform]{\includegraphics[width=0.15\textwidth]{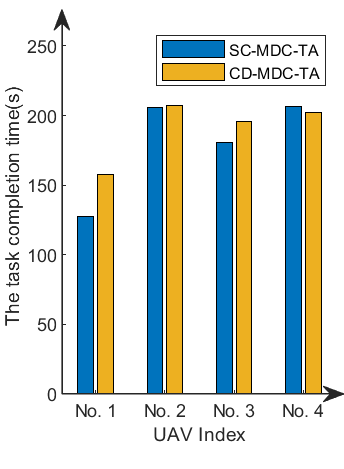}\label{212-2-time}}
	\subfloat[Clustered]{\includegraphics[width=0.15\textwidth]{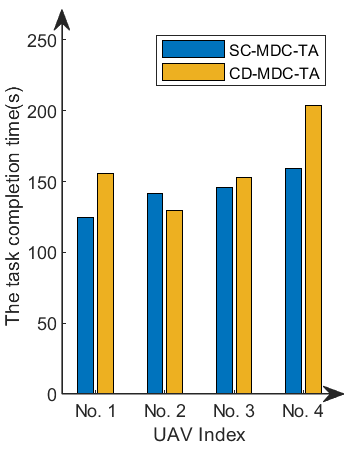}\label{213-2-time}} 	
	\caption{The comparison of task completion time under different task distributions.}\label{taskstime}
\end{figure}

\begin{figure}
	\centering
	\subfloat[Randomn]{\includegraphics[width=0.15\textwidth]{figures/212-3-energy}\label{211-3-energy}} 	
	\subfloat[Uniform]{\includegraphics[width=0.15\textwidth]{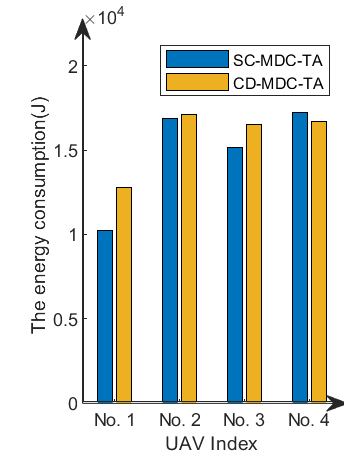}\label{212-3-energy}}
	\subfloat[Clustered]{\includegraphics[width=0.15\textwidth]{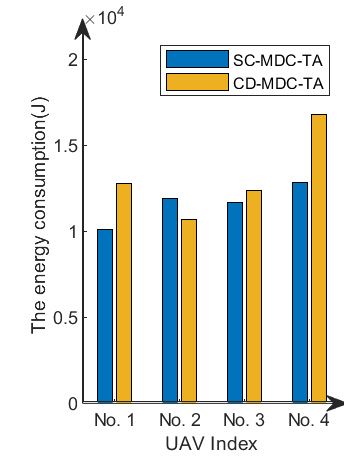}\label{213-3-energy}} 	
	\caption{The comparison of task completion time under different task distributions.}\label{tasksenergy}
\end{figure}

\begin{figure}
	\centering
	\subfloat[Task completion time]{\includegraphics[width=0.23\textwidth]{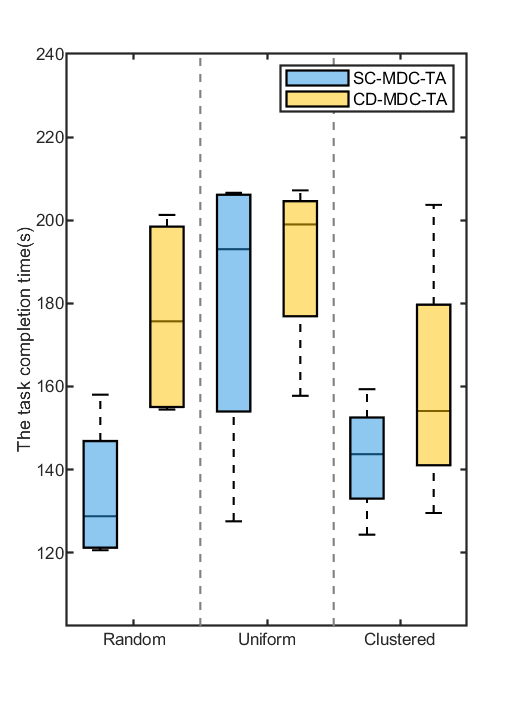}\label{boxtime-time-time}} 	
	\subfloat[UAV energy consumption]{\includegraphics[width=0.23\textwidth]{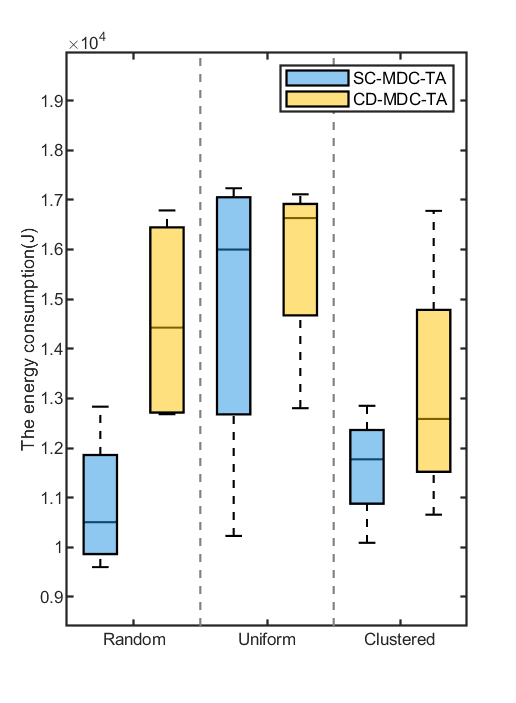}\label{boxtime-time-energy}}	
	\caption{Performance boxplot comparison under different task distributions.}\label{taskscompare}
\end{figure}

\subsection{Performance Comparison From Different Take-off Positions}

        For the diverse demands of maritime tasks, the selection of UAV take-off positions is crucial. Center theoretically minimize average response times, while corners reduce round-trip durations for specific zones. Edge leverage natural factors like wind direction and currents, or facilitate swift responses and evacuations in emergencies. Fig. \ref{locationsallocation} illustrates the impact of take-off positions on MDC task allocation. The proposed SC-MDC-TA algorithm, with its intelligent adjustments, accommodates UAVs from varied starting positions.

        Fig. \ref{locationtime} and Fig. \ref{locationenergy} compare the task completion time and energy consumption of the proposed SC-MDC-TA and CD-MDC-TA algorithms at different UAV take-off positions. Regardless of whether the take-off is at the center, corners, or edge of the sea area, the proposed SC-MDC-TA algorithm consistently shows a decisive advantage. Notably, compared to the benchmark, the proposed SC-MDC-TA algorithm achieves time savings of 15.2\%, 24.2\%, and 5.3\% at the center, corners, and edge, respectively, while reducing energy consumption by 16.0\%, 25.5\%, and 5.6\%, respectively. Additionally, Fig. \ref{locationcompare} visually reveals the performance distribution comparison of the two algorithms in terms of task completion time and UAV energy consumption through box plots. The compactness of the data distribution in the proposed SC-MDC-TA algorithm highlights its robustness and consistency in handling different UAV take-off positions. The lower variance further confirms the algorithm's strong control capabilities, effectively mitigating performance instability caused by UAV take-off positions, and providing a solid guarantee for reliable services in the maritime domain.

        The proposed SC-MDC-TA algorithm performs best when UAVs take off from the corners, followed by the center, and performs weakest when taking off from the edge, mainly due to the following reasons. Firstly, taking off from the corners maximizes the coverage of UAVs, minimizes flight distances, and improves the efficiency of task allocation. Secondly, although taking off from the center facilitates rapid responses in all directions, it may lead to increased energy consumption when performing tasks far from the launch point. Finally, taking off from the edge limits the coverage of UAVs, resulting in higher task response times and energy consumption. Therefore, when UAVs take off from the corners, the proposed SC-MDC-TA algorithm most effectively balances task execution efficiency and energy management, achieving optimal performance.

\begin{figure*}
	\centering
	\subfloat[2 UAVs]{\includegraphics[width=0.33\textwidth]{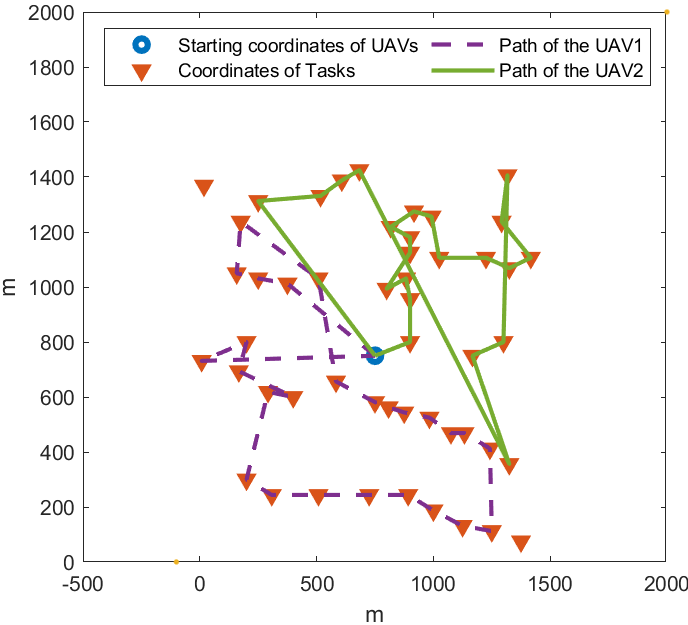}\label{211-1-A}} 	
	\subfloat[3 UAVs]{\includegraphics[width=0.33\textwidth]{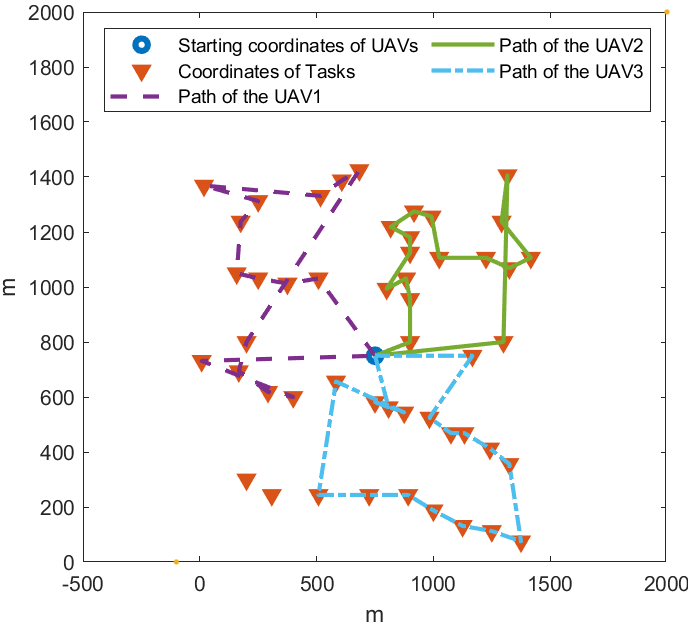}\label{211-1-A}}
	\subfloat[4 UAVs]{\includegraphics[width=0.33\textwidth]{figures/112-1-A}\label{211-1-A}}
	\caption{The MDC task allocation based on the proposed SC-MDC-TA algorithm under different UAV numbers.}\label{UAVallocation}
\end{figure*}

\begin{figure}
	\centering
	\subfloat[2 UAVs]{\includegraphics[width=0.15\textwidth]{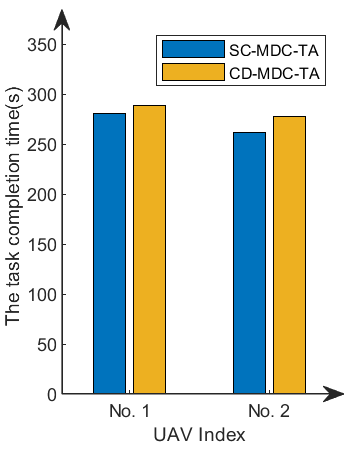}\label{211-2-time}} 	
	\subfloat[3 UAVs]{\includegraphics[width=0.15\textwidth]{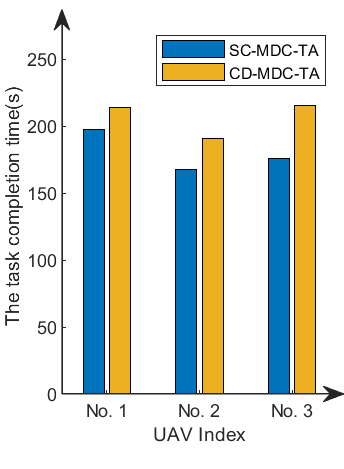}\label{211-2-time}} 
	\subfloat[4 UAVs]{\includegraphics[width=0.15\textwidth]{figures/112-2-time}\label{211-2-time}} 	
	\caption{The comparison of task completion time under different UAV numbers.}\label{UAVtime}
\end{figure}

\begin{figure}
	\centering
	\subfloat[2 UAVs]{\includegraphics[width=0.15\textwidth]{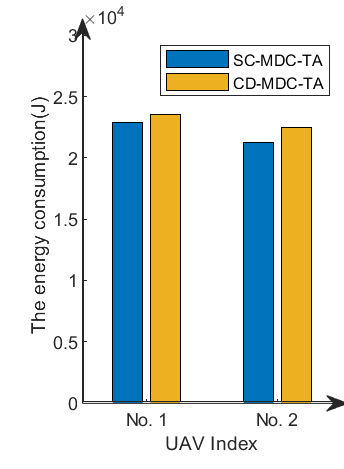}\label{211-3-energy}} 	
	\subfloat[3 UAVs]{\includegraphics[width=0.15\textwidth]{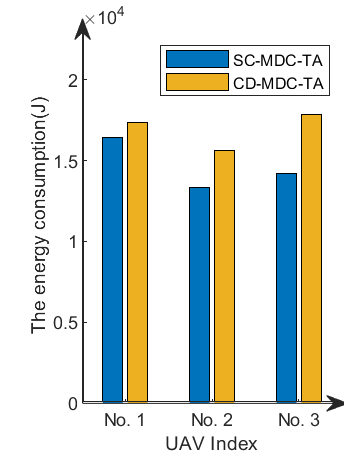}\label{211-3-energy}} 
	\subfloat[4 UAVs]{\includegraphics[width=0.15\textwidth]{figures/112-3-energy}\label{211-3-energy}} 	
	\caption{The comparison of task completion time under different UAV numbers.}\label{UAVenergy}
\end{figure}

\begin{figure}
	\centering
	\subfloat[Task completion time]{\includegraphics[width=0.23\textwidth]{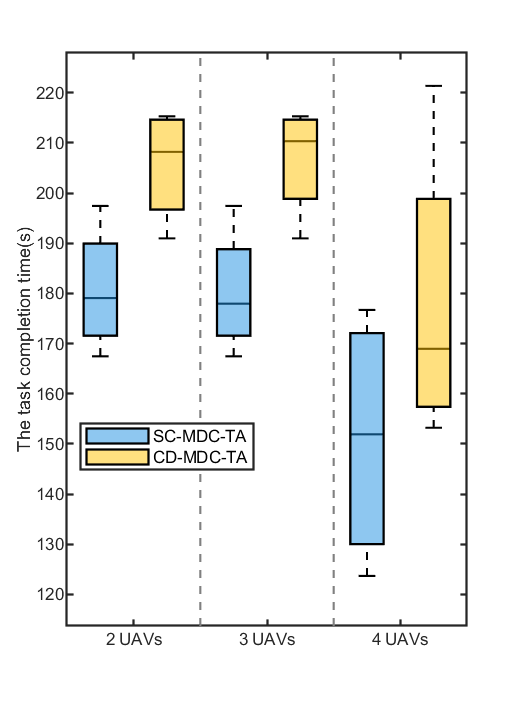}\label{boxtime-time-time}} 	
	\subfloat[UAV energy consumption]{\includegraphics[width=0.23\textwidth]{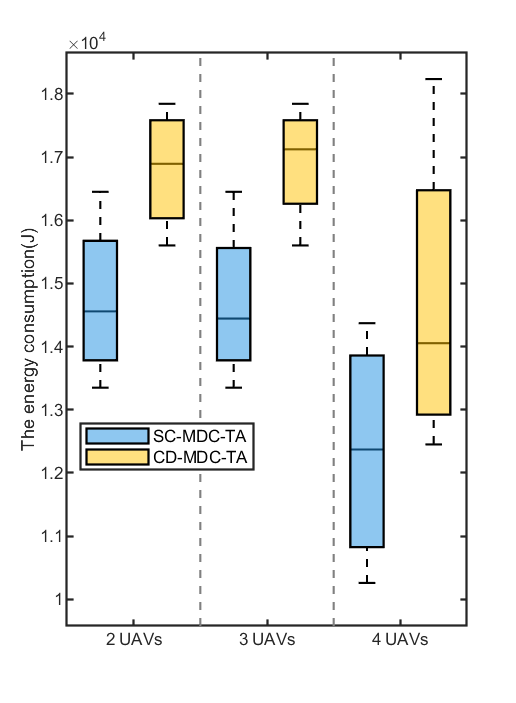}\label{boxtime-time-energy}}	
	\caption{Performance boxplot comparison under different UAV numbers.}\label{UAVcompare}
\end{figure}

\subsection{Performance Comparison Under Different Task Distributions}

        Addressing the diverse distribution characteristics inherent in MDC tasks completed by multiple UAVs, we conduct an in-depth analysis of the impacts of various task distributions. A random distribution implies high uncertainty in task positions, necessitating algorithms with rapid response capabilities and adaptable planning strategies. Uniform distribution, on the other hand, requires algorithms capable of evenly covering all areas to ensure no task is overlooked. Clustered distribution emphasizes the effective concentration of resources in high-density task zones and the optimization of paths. Fig. \ref{tasksallocation} significantly underscores the decisive role of task allocation patterns in MDC task assignments, highlighting the imperative for algorithm designs to possess a high degree of adaptability and flexibility towards diverse task distributions. This is crucial to maintain superior performance across any deployment scenario.

        The comparison between Fig. \ref{taskstime} and Fig. \ref{tasksenergy} clearly illustrates the performance gap between the algorithms under different task distribution patterns. Regardless of whether the tasks are random, uniform, or clustered distribution, the SC-MDC-TA algorithm consistently maintains an advantage. Specifically, compared to the CD-MDC-TA algorithm, the SC-MDC-TA algorithm achieves reductions in task completion time of 24.2\%, 5.6\%, and 11.0\% under random, uniform, and clustered distributions, respectively, while energy consumption is reduced by 25.5\%, 5.9\%, and 11.6\%, respectively. Further, Fig. \ref{taskscompare}, through box plots, presents the stability of each algorithm's performance across various task distributions. The compact data distribution of SC-MDC-TA highlights its stability and reliability when handling diverse task layouts, reinforcing its capability to swiftly adapt to different MDC task distributions and ensuring efficient task execution.

        The proposed SC-MDC-TA algorithm performs best under random task distribution due to its dynamic adaptability. It can fully utilize its ability to dynamically adjust, quickly optimizing flight paths based on real-time task demands, reducing unnecessary flights, and thus improving task completion speed and energy efficiency. Facing clustered task distributions, it can concentrate resources, focus on high-density areas, optimize paths, maximize local efficiency, and enhance overall task completion quality. In contrast, the uniform task distribution model poses higher demands on UAV flight paths, increasing flight distances and potentially leading to greater energy consumption. Therefore, the superiority of the proposed SC-MDC-TA algorithm is more pronounced in situations of random and clustered task distributions, effectively achieving MDC tasks.

\subsection{Performance Comparison Under Different UAV Numbers}

        We conduct an analysis on the execution of MDC tasks with varying UAV numbers. When only two UAVs are assigned to the task, resources are limited, requiring comprehensive coverage of the task area to ensure the smooth execution of the task. As the UAV number increases to three, the system confronts more intricate challenges in resource coordination and path optimization, where the algorithm demonstrates good scalability. Upon further escalation to a fleet of four UAVs, the algorithm optimally utilizes the available resources, enabling heightened levels of UAV collaboration for the fulfillment of MDC tasks. Fig. \ref{UAVallocation} illustrates the impact of the UAV numbers on the allocation of MDC tasks, highlighting how the algorithm is adaptable to the diverse requirements of different UAV fleet sizes. This adaptability and potential for optimization in the face of varied real-world conditions underscore the practical efficacy and versatility of the algorithm.

        The stark contrasts in Fig. \ref{UAVtime} and Fig. \ref{UAVenergy} highlight the performance differences between the algorithms under various UAV numbers. Specifically, compared to the benchmark, the proposed SC-MDC-TA algorithm achieves a reduction in task completion time of 4.2\%, -4.6\%, and 15.2\%, and a decrease in energy consumption of 4.2\%, -4.9\%, and 16.0\% when the UAV number is 2, 3, and 4, respectively. Moreover, Fig. \ref{UAVcompare} presents box plots that vividly illustrate these comparisons. The tight clustering of data points for the SC-MDC-TA algorithm signifies its superior consistency and reliability across differing quantities of UAVs. These findings validate the adaptability of the SC-MDC-TA algorithm to fluctuations in the UAV numbers, ensuring relatively efficient task execution. Notably, the best performance is observed with four UAVs; second-best with two UAVs, whereas efficiency dips slightly with a fleet of three UAVs.
	
		The differences under various UAV numbers reveal its balance between resource scheduling and task execution. Wit four UAVs, the proposed SC-MDC-TA algorithm exhibits synergy. Through the MDC task allocation, the proposed SC-MDC-TA algorithm not only shortens the task execution time but also reduces energy consumption. The intelligent scheduling and efficient collaboration of the UAV fleet amplify overall efficiency. Despite the limited number, two UAVs can still maintain a high level of task execution efficiency and energy-saving effects thanks to the proposed SC-MDC-TA algorithm's path optimization and strategic adjustments. When the UAV number is three, the algorithm faces a trade-off between resource allocation and efficiency improvements, affecting the smoothness of task execution and leading to additional flights and energy waste. In this case, the proposed SC-MDC-TA algorithm needs to make compromises in task segmentation and route planning, thus affecting overall performance.
	
\section{Conclusions}
	
        In this work, we have focused on the optimization problem of UAV-assisted MDC task allocation, considering comprehensively the spatial and temporal requirements of MDC tasks and the mobility of UAVs. We have introduced an SC-based MDC network model, treating UAVs as workers, with the maritime work station responsible for allocating data collection and data return tasks to them. Additionally, we have proposed the SC-MDC-TA algorithm, which effectively enhances the holistic performance through the quality estimation algorithm and the reverse auction algorithm. Combining comprehensive comparisons of the task completion time and UAV energy consumption, the proposed SC-MDC-TA algorithm has flexibly adjusted the MDC task allocation results to meet maritime service scenarios. Simulation results have indicated that, compared to the benchmark, the proposed SC-MDC-TA algorithm has significantly reduced the task completion time and decreased the UAV energy consumption. This work has validated the applicability of the proposed SC-MDC-TA algorithm in diverse maritime service scenarios, demonstrating its superior and stable performance, making it an ideal choice for the MDC task allocation.

  \bibliographystyle{IEEEtrans}   
  \bibliography{refSCMDC}

\begin{thebibliography}{10}
\providecommand{\url}[1]{#1}
\csname url@samestyle\endcsname
\providecommand{\newblock}{\relax}
\providecommand{\bibinfo}[2]{#2}
\providecommand{\BIBentrySTDinterwordspacing}{\spaceskip=0pt\relax}
\providecommand{\BIBentryALTinterwordstretchfactor}{4}
\providecommand{\BIBentryALTinterwordspacing}{\spaceskip=\fontdimen2\font plus
\BIBentryALTinterwordstretchfactor\fontdimen3\font minus
  \fontdimen4\font\relax}
\providecommand{\BIBforeignlanguage}[2]{{%
\expandafter\ifx\csname l@#1\endcsname\relax
\typeout{** WARNING: IEEEtran.bst: No hyphenation pattern has been}%
\typeout{** loaded for the language `#1'. Using the pattern for}%
\typeout{** the default language instead.}%
\else
\language=\csname l@#1\endcsname
\fi
#2}}
\providecommand{\BIBdecl}{\relax}
\BIBdecl

\bibitem{1}
T.~Wei, W.~Feng, Y.~Chen, C.-X. Wang, N.~Ge, and J.~Lu, ``Hybrid
  satellite-terrestrial communication networks for the maritime internet of
  things: Key technologies, opportunities, and challenges,'' \emph{IEEE
  Internet of Things Journal}, vol.~8, no.~11, pp. 8910--8934, 2021.

\bibitem{2}
T.~Xia, M.~M. Wang, J.~Zhang, and L.~Wang, ``Maritime internet of things:
  Challenges and solutions,'' \emph{IEEE Wireless Communications}, vol.~27,
  no.~2, pp. 188--196, 2020.

\bibitem{3}
F.~Bekkadal, ``Future maritime communications technologies,'' in \emph{OCEANS
  2009-EUROPE}, 2009, pp. 1--6.

\bibitem{4}
B.~Lin, L.~Zhao, H.~A. Suraweera, T.~H. Luan, D.~Niyato, and D.~T. Hoang,
  ``Guest editorial special issue on internet of things for smart ocean,''
  \emph{IEEE Internet of Things Journal}, vol.~7, no.~10, pp. 9675--9677, 2020.

\bibitem{5}
C.~Yu, J.~Li, C.~Zhang, H.~Li, R.~He, and B.~Lin, ``Maritime broadband
  communications: Applications, challenges and an offshore 5g-virtual mimo
  paradigm,'' in \emph{2020 IEEE Intl Conf on Parallel \& Distributed
  Processing with Applications, Big Data \& Cloud Computing, Sustainable
  Computing \& Communications, Social Computing \& Networking
  (ISPA/BDCloud/SocialCom/SustainCom)}, 2020, pp. 1286--1291.

\bibitem{6}
T.~Yang, H.~Feng, C.~Yang, Y.~Wang, J.~Dong, and M.~Xia, ``Multivessel
  computation offloading in maritime mobile edge computing network,''
  \emph{IEEE Internet of Things Journal}, vol.~6, no.~3, pp. 4063--4073, 2019.

\bibitem{7}
Y.~Li, Y.~Zhang, W.~Li, and T.~Jiang, ``Marine wireless big data: Efficient
  transmission, related applications, and challenges,'' \emph{IEEE Wireless
  Communications}, vol.~25, no.~1, pp. 19--25, 2018.

\bibitem{8}
J.~Tiemann, O.~Feldmeier, and C.~Wietfeld, ``Supporting maritime search and
  rescue missions through uas-based wireless localization,'' in \emph{2018 IEEE
  Globecom Workshops (GC Wkshps)}, 2018, pp. 1--6.

\bibitem{9}
T.~Yang, H.~Liang, N.~Cheng, R.~Deng, and X.~Shen, ``Efficient scheduling for
  video transmissions in maritime wireless communication networks,'' \emph{IEEE
  Transactions on Vehicular Technology}, vol.~64, no.~9, pp. 4215--4229, 2015.

\bibitem{10}
L.~Lyu, Z.~Chu, B.~Lin, Y.~Dai, and N.~Cheng, ``Fast trajectory planning for
  uav-enabled maritime iot systems: A fermat-point based approach,'' \emph{IEEE
  Wireless Communications Letters}, vol.~11, no.~2, pp. 328--332, 2022.

\bibitem{11}
X.~Li, W.~Feng, Y.~Chen, C.-X. Wang, and N.~Ge, ``Maritime coverage enhancement
  using uavs coordinated with hybrid satellite-terrestrial networks,''
  \emph{IEEE Transactions on Communications}, vol.~68, no.~4, pp. 2355--2369,
  2020.

\bibitem{12}
N.~E. Leonard, D.~A. Paley, F.~Lekien, R.~Sepulchre, D.~M. Fratantoni, and
  R.~E. Davis, ``Collective motion, sensor networks, and ocean sampling,''
  \emph{Proceedings of the IEEE}, vol.~95, no.~1, pp. 48--74, 2007.

\bibitem{41}
L.~Shen, N.~Wang, Z.~Zhu, Y.~Fan, X.~Ji, and X.~Mu, ``Uav-enabled data
  collection for mmtc networks: Aem modeling and energy-efficient trajectory
  design,'' in \emph{ICC 2020 - 2020 IEEE International Conference on
  Communications (ICC)}, 2020, pp. 1--6.

\bibitem{29}
L.~Wang, Z.~Yu, Q.~Han, B.~Guo, and H.~Xiong, ``Multi-objective optimization
  based allocation of heterogeneous spatial crowdsourcing tasks,'' \emph{IEEE
  Transactions on Mobile Computing}, vol.~17, no.~7, pp. 1637--1650, 2018.

\bibitem{30}
B.~Guo, Y.~Liu, W.~Wu, Z.~Yu, and Q.~Han, ``Activecrowd: A framework for
  optimized multitask allocation in mobile crowdsensing systems,'' \emph{IEEE
  Transactions on Human-Machine Systems}, vol.~47, no.~3, pp. 392--403, 2017.

\bibitem{31}
P.~Cheng, X.~Lian, L.~Chen, J.~Han, and J.~Zhao, ``Task assignment on
  multi-skill oriented spatial crowdsourcing,'' \emph{IEEE Transactions on
  Knowledge and Data Engineering}, vol.~28, no.~8, pp. 2201--2215, 2016.

\bibitem{13}
B.~Guo, Y.~Liu, L.~Wang, V.~O.~K. Li, J.~C.~K. Lam, and Z.~Yu, ``Task
  allocation in spatial crowdsourcing: Current state and future directions,''
  \emph{IEEE Internet of Things Journal}, vol.~5, no.~3, pp. 1749--1764, 2018.

\bibitem{14}
H.~Xiong, D.~Zhang, G.~Chen, L.~Wang, V.~Gauthier, and L.~E. Barnes, ``icrowd:
  Near-optimal task allocation for piggyback crowdsensing,'' \emph{IEEE
  Transactions on Mobile Computing}, vol.~15, no.~8, pp. 2010--2022, 2016.

\bibitem{20}
L.~Wang, Z.~Yu, Q.~Han, D.~Yang, S.~Pan, Y.~Yao, and D.~Zhang, ``Compact
  scheduling for task graph oriented mobile crowdsourcing,'' \emph{IEEE
  Transactions on Mobile Computing}, vol.~21, no.~7, pp. 2358--2371, 2022.

\bibitem{21}
S.~S. Bhatti, J.~Fan, K.~Wang, X.~Gao, F.~Wu, and G.~Chen, ``An approximation
  algorithm for bounded task assignment problem in spatial crowdsourcing,''
  \emph{IEEE Transactions on Mobile Computing}, vol.~20, no.~8, pp. 2536--2549,
  2021.

\bibitem{22}
Y.~Jiao, P.~Wang, D.~Niyato, B.~Lin, and D.~I. Kim, ``Mechanism design for
  wireless powered spatial crowdsourcing networks,'' \emph{IEEE Transactions on
  Vehicular Technology}, vol.~69, no.~1, pp. 920--934, 2020.

\bibitem{15}
S.~Yang, F.~Wu, S.~Tang, X.~Gao, B.~Yang, and G.~Chen, ``On designing data
  quality-aware truth estimation and surplus sharing method for mobile
  crowdsensing,'' \emph{IEEE Journal on Selected Areas in Communications},
  vol.~35, no.~4, pp. 832--847, 2017.

\bibitem{16}
Z.~Zheng, F.~Wu, X.~Gao, H.~Zhu, S.~Tang, and G.~Chen, ``A budget feasible
  incentive mechanism for weighted coverage maximization in mobile
  crowdsensing,'' \emph{IEEE Transactions on Mobile Computing}, vol.~16, no.~9,
  pp. 2392--2407, 2017.

\bibitem{18}
H.~Wang, E.~Wang, Y.~Yang, J.~Wu, and F.~Dressler, ``Privacy-preserving online
  task assignment in spatial crowdsourcing: A graph-based approach,'' in
  \emph{IEEE INFOCOM 2022 - IEEE Conference on Computer Communications}, 2022,
  pp. 570--579.

\bibitem{17}
X.~Chen, L.~Zhang, Y.~Pang, B.~Lin, and Y.~Fang, ``Timeliness-aware incentive
  mechanism for vehicular crowdsourcing in smart cities,'' \emph{IEEE
  Transactions on Mobile Computing}, vol.~21, no.~9, pp. 3373--3387, 2022.

\bibitem{19}
S.~Wu, Y.~Wang, and X.~Tong, ``Multi-objective task assignment for maximizing
  social welfare in spatio-temporal crowdsourcing,'' \emph{China
  Communications}, vol.~18, no.~11, pp. 11--25, 2021.

\bibitem{24}
J.~Qin, Q.~Ma, Y.~Shi, and L.~Wang, ``Recent advances in consensus of
  multi-agent systems: A brief survey,'' \emph{IEEE Transactions on Industrial
  Electronics}, vol.~64, no.~6, pp. 4972--4983, 2017.

\bibitem{26}
Y.~Cao, W.~Yu, W.~Ren, and G.~Chen, ``An overview of recent progress in the
  study of distributed multi-agent coordination,'' \emph{IEEE Transactions on
  Industrial Informatics}, vol.~9, no.~1, pp. 427--438, 2013.

\bibitem{25}
D.~Kidston and T.~Kunz, ``Challenges and opportunities in managing maritime
  networks,'' \emph{IEEE Communications Magazine}, vol.~46, no.~10, pp.
  162--168, 2008.

\bibitem{27}
Y.~Zhao and Q.~Han, ``Spatial crowdsourcing: current state and future
  directions,'' \emph{IEEE Communications Magazine}, vol.~54, no.~7, pp.
  102--107, 2016.

\bibitem{zengyong1}
Y.~Zeng, Q.~Wu, and R.~Zhang, ``Accessing from the sky: A tutorial on uav
  communications for 5g and beyond,'' \emph{Proceedings of the IEEE}, vol. 107,
  no.~12, pp. 2327--2375, 2019.

\bibitem{zengyong2}
Y.~Zeng and R.~Zhang, ``Energy-efficient uav communication with trajectory
  optimization,'' \emph{IEEE Transactions on Wireless Communications}, vol.~16,
  no.~6, pp. 3747--3760, 2017.

\bibitem{zengyong3}
Y.~Zeng, J.~Xu, and R.~Zhang, ``Energy minimization for wireless communication
  with rotary-wing uav,'' \emph{IEEE Transactions on Wireless Communications},
  vol.~18, no.~4, pp. 2329--2345, 2019.

\bibitem{bowen1}
Z.~Na, B.~Li, X.~Liu, J.~Wang, M.~Zhang, Y.~Liu, and B.~Mao, ``Uav-based
  wide-area internet of things: An integrated deployment architecture,''
  \emph{IEEE Network}, vol.~35, no.~5, pp. 122--128, 2021.

\bibitem{bowen2}
B.~Li, Z.~Na, and B.~Lin, ``Uav trajectory planning from a comprehensive energy
  efficiency perspective in harsh environments,'' \emph{IEEE Network}, vol.~36,
  no.~4, pp. 62--68, 2022.

\end{thebibliography}

\end{document}